\documentclass{aa}
\usepackage{txfonts}
\usepackage{graphicx}
\newcommand{\Msun}{\mbox{M}_{\sun}}
\newcommand{\Mpc}{$h^{-1}$Mpc }
\begin{document}

\title{Environments of Nearby Quasars in Sloan Digital Sky Survey}

\author{H. Lietzen \inst{1} 
  \and P. Hein\"am\"aki\inst{1}
  \and P. Nurmi \inst{1}
  \and E. Tago \inst{2}
  \and E. Saar \inst{2}
  \and J. Liivam\"agi \inst{2}
  \and E. Tempel \inst{2}
  \and M. Einasto \inst{2}
  \and J. Einasto \inst{2}
  \and M. Gramann \inst{2}
  \and L. O. Takalo \inst{1}
  }
\institute{Tuorla Observatory, Department of Physics and Astronomy, 
University of Turku, V\"{a}is\"{a}l\"{a}ntie 20, FI-21500 Piikki\"{o}, Finland
\and Tartu Observatory, 61602 T\~{o}ravere, Tartumaa, Estonia 
}

\date{Received / Accepted }

\abstract
{For the first time spectroscopic galaxy redshift surveys are reaching the
scales where galaxies can be studied together with
the nearest quasars. This gives an opportunity to study the dependence between
the activity of a quasar and its environment in a more extensive way than 
before.}
{We study the spatial distribution of galaxies and groups of galaxies in the 
environments of low redshift quasars in the Sloan Digital Sky Survey (SDSS).
Our aim is to understand how the nearby quasars are embedded in the local and 
global density field of galaxies and how the environment affects quasar 
activity.}
{We analyse the environments of nearby quasars using number counts of galaxies.
We also study the dependence of group properties 
to their distance to the nearest quasar. The large scale environments are 
studied by analysing the locations of quasars in the luminosity density field.}
{ Our study of the number counts of galaxies in quasar environments 
shows an underdensity of bright galaxies at a few Mpc from  
quasars. Also, the groups of galaxies that have a quasar closer
than 2Mpc are poorer and less luminous than in 
average. Our analysis on the luminosity density field shows that quasars
clearly avoid rich superclusters. Nearby quasars seem to be located in 
outskirts of superclusters or in filaments connecting them.}
{Our results suggest that quasar evolution may be affected by density 
variations both
on supercluster scales and in the local environment.}

\maketitle

\section{Introduction}

Active galactic nuclei (AGN) are believed to have evolved by mergers between 
galaxies that cause 
inflow of gas which fuels the growth of the black hole. According to the 
theory of quasar formation by Hopkins et al. (\cite{Hopkins2005b}), 
there are frequent mergers 
between galaxies during the hierarcial growth of structure. Mergers that
involve gas-rich galaxy progenitors produce inflows of gas that cause 
starbursts. High gas densities fuel the black hole growth, and the galaxy is 
seen as a luminous infrared galaxy (LIRG). As the black hole mass increases, 
the feedback energy starts to expel the gas fueling accretion. The galaxy is 
then seen as an optical quasar. Finally the feedback of the active galactic 
nucleus terminates further 
black hole growth, after which the galaxy becomes an ordinary galaxy with a 
dead quasar.
Although this is the prevailing paradigm for quasar evolution, 
it is also possible that it may not be the only mechanism for initiating 
activity for all types of active galaxies. Activity may also be triggered 
by secular processes, such as bar instabilities 
(Sellwood \& Moore \cite{Sellwood}.)

One way of testing different theories for nuclear activity is to 
study the environments of AGN. If mergers are important to quasar activity, 
the environments of quasars should allow mergers to happen. 
In clusters that have a high 
velocity dispersion galaxies do not have enough time to merge, and 
therefore these should not be as likely locations for quasars as small groups of 
galaxies that usually have lower velocity dispersions 
(Hopkins et al. \cite{Hopkins2005b}).

It is well known that environmental density affects properties of 
galaxies; an example of that is the morphology--density relation (Dressler \cite{Dressler}).
This dependency has been studied, e.g., by Kauffmann et al. (\cite{Kauffmann}). 
Their results show that the stellar mass of galaxies increases and their star 
formation rate decreases
with the growth of the local density. They also find that a larger 
fraction of galaxies host AGN in low-density environments.
It is obvious that morphologies, colors, and star formation rates of galaxies 
are related 
to the assembly history of their halos (Maulbetsch et al. \cite{Maul}).

Another reason to study the environments of quasars is to test the unification
schemes of different types of
objects (Coldwell \& Lambas \cite{Coldwell}).
Different types of AGN
are thought to be physically similar objects that are viewed from 
different directions (Antonucci \cite{Antonucci}). If this is true, all
AGN should have similar environments since the properties of the 
environments do not depend on the viewing angle.
It is also possible that mergers are more important to the evolution of 
some types of AGN while others might be triggered through other mechanisms.
Activity and luminosity of the AGN may play an important role that give clues 
for their merger history and associated environment.

Earlier studies of the AGN environments have 
mostly used galaxy counts in the fields of individual objects.
The first results were from Bahcall et al. (\cite{Bahcall}) who showed that 
five quasars lie within the approximate geometrical boundaries of clusters of 
galaxies. Later much work has been done to estimate the spatial 
galaxy--galaxy cross-correlation amplitude; this was initiated by Longair \& 
Seldner (\cite{Longair}). 
The correlation studies have shown that quasars mostly occupy 
enhanced local densities, such as groups or clusters of galaxies.
For example, De Robertis et al. (\cite{DeRobertis}) 
found that the environments of Seyfert 2 type galaxies are not different from 
the control sample of normal galaxies, while the environments of Seyfert 1 
galaxies are clearly poorer. The environments of BL Lac objects have been 
studied by Smith et al. (\cite{Smith}) and 
Wurtz et al. (\cite{Wurtz}). Both  
these studies show that BL Lac objects lie in poor environments, usually in 
clusters of the Abell richness class 0 or less. According to McLure \& Dunlop 
(\cite{McLure}) radio galaxies are distributed quite evenly over all cluster richness 
levels, but quasars are more frequent in poor environments. This was also 
found by Wold et al. (\cite{Wold}); their results show that although there are 
a few quasars in rich environments, most of them seem to prefer galaxy 
groups or clusters of the Abell richness class 0. As a summary, 
most AGN lie in very poor clusters, with only a few more companion 
galaxies than are found in the background field, for a few hundred kpc scale.

The Sloan Digital Sky Survey (SDSS) offers new possibilities to study the large-scale structure
of the Universe. As concerns the study of the environments of quasars, SDSS contains 
not only data for detailed studies of individual targets, but  
the large number of quasars and galaxies makes it possible to study these 
environments on larger scales than before.

Li et al. (\cite{Li}) have measured AGN--galaxy 
cross-correlations for a large sample of quasars 
and galaxies in the SDSS Data Release 4. They find that on the scales between 
100 kpc and 1 Mpc AGN hosts are clustered more weakly than inactive galaxies, 
while at smaller scales their clustering is slightly stronger. Their explanation
for this result is that AGN reside preferentially at the centers of dark 
matter halos.

A slightly different approach was used by 
Miller et al. (\cite{Miller}), who studied the fraction of galaxies that possess 
an AGN, as a function of the environment. They used galaxies in the redshift range 
$0.05 \geq z \geq 0.095$ with the absolute magnitude $M<-20.0$ from the Early Data Release 
of the SDSS. Their main result is that $\sim 40$\,\% of galaxies in their 
sample have an AGN, and they do not find any relation between the AGN fraction 
and the local galaxy density.

One of the most extensive studies of the environments of  
nearby quasars was done by Coldwell \& Lambas (\cite{Coldwell}). 
They used quasars and galaxies at redshifts less than 0.2 in the third 
Data Release of the SDSS in order to study the properties of galaxies 
in the environments of quasars. Their main results are that  nearby 
quasars avoid high density regions, and that the galaxies close to quasars 
usually have a disc-type morphology and a high star formation rate. 

As the data volume provided by the SDSS has increased,
environments of quasars have become an increasingly popular subject to study.
One of the latest studies was done by Strand et al. (\cite{Strand}), who compared 
the environments of quasars to the AGN chosen from the photometric galaxy sample of 
the SDSS Data Release 5. The galaxy densities were measured from the photometric 
catalog that contains much more galaxies than the spectroscopic sample, but the 
lower accuracy of redshifts limits the analysis. Their main result is that 
quasars lie in higher density regions than the AGN at scales of 
less than 2 Mpc.

In this paper we extend the previous studies to the galaxy group and supercluster 
scales. We study the environments of quasars at three different scales. First 
we concentrate on galaxies around quasars, analysing the number density 
of galaxies at different distances from quasars. At the next scale, we 
study the relations 
between the group properties and the distance of its closest quasar. For the 
largest scale in this study, we use the luminosity density fields, 
which we have constructed from the galaxy catalogs. 
This allows us to describe the locations of quasars at supercluster scales.
Throughout this work we assume $\Omega_m=0.3$, $\Omega_\Lambda=0.7$, and 
$H_0=100h$ km s$^{-1}$ Mpc$^{-1}$. Absolute magnitudes are given for $h=1$.

\section{Data}

In this paper we use the spectroscopic catalogs of quasars and galaxies 
from the fifth (DR5) and sixth (DR6) Data Releases of the SDSS 
(Adelman-McCarthy et al. \cite{Adelman}). 
The original SDSS-DR5 galaxy sample contains 674749 galaxies, and 
the original quasar sample contains 90611 quasars. The DR6 contains 790,220
galaxies. We use three different samples: volume limited samples of galaxies 
from the DR5
for estimating the number densities of galaxies, volume limited samples of 
groups of galaxies from the DR6 for studying the group properties, and 
a magnitude limited sample of galaxies for studying the supercluster scales.

The final data for galaxies in this paper comes from the 
Tago group catalog (Tago et al. \cite{Tago}) that is based on the
main galaxy sample obtained from the SDSS Data Archive Server, with 488725 
galaxies. Applying the lower ($z=0.009$) and upper ($z=0.12$) redshift limits 
and excluding several duplicate galaxies in the original data 
the number of galaxies was reduced to 387063. 

To study how nearby quasars are located among galaxies 
we extracted four volume 
limited subsamples of galaxies with the absolute magnitude limits of 
$M_{\mathrm{gal}}<-21.0$,
$M_{\mathrm{gal}}<-21.5$, $M_{\mathrm{gal}}<-22.0$, $M_{\mathrm{gal}}<-22.5$ (Petrosian r-magnitudes). 
The 
samples contain 106378, 37181, 7790 and 985 galaxies, respectively. For
the limit $M_{\mathrm{gal}}<-21.0$ the co-moving distance is chosen to be less than
500\,\Mpc, while the lack of very nearby quasars sets a lower distance limit 
of 200\,\Mpc. 
The value for $M^*(r)$ for galaxies in the SDSS is $-20.44$, and 
therefore our samples consist of the brightest end of the luminosity function 
(Blanton et al. \cite{Blanton}).

The final group catalog we use for our group analysis is a volume limited 
sample based on the DR6 with the magnitudes 
$-22.0 < M_{\mathrm{group}}<-21.0$. 
This sample contains 9581 groups, and extends up to the co-moving 
distance of 450 $h^{-1}$Mpc.
Groups in the SDSS DR6 have been defined applying the well-known
Friends-of-Friends cluster analysis method introduced by Turner \& Gott 
(\cite{Turner}), and modified by E. Saar. In this algorithm galaxies have 
been linked into systems using a valiable linking length. By this definition 
a galaxy belongs to a group of galaxies if this galaxy has at least one 
group member galaxy closer than the linking length. In a flux-limited 
subsample the linking length was increased along with distance in order to 
keep virial properties, such as virial radius, maximum size in the sky 
projection and velocity dispersion similar at all distances. In the volume 
limited subsample that was used in this study, the linking length was 
constant inside one particular subsample, but was changed from subsample to 
subsample of various absolute magnitude limits in order to keep the virial 
properties similar in all volume limited subsamples.

Finally we use the Tago et al. (\cite{Tago}) magnitude limited sample of galaxies  
for determining the luminosity 
density field. The limiting magnitudes of the complete 
spectroscopic sample of the SDSS catalog used here are 14.5 and 17.77 in 
r band. See Tago et al. (\cite {Tago}) for the description of the selection effects 
taken into account for this magnitude-limited galaxy sample.

As our final quasar data we use the value added quasar catalog based on the DR5 
(Schneider et al. \cite{Schneider}), which contains 77429 quasars. 
These quasars have the absolute $i$ band magnitude $M_{\mathrm{i}}<-22.0$, 
and have emission lines with line widths larger than 1000 km/s.
With these conditions the sample should contain only classical quasars, while
the original SDSS quasar catalog contains also other kinds of active galaxies.
Due to the limits of our galaxy and group catalogs we include only quasars at the
redshifts $0.078<z<0.172$. We also require the quasars to satisfy the target 
selection flag BEST, which marks the results obtained using the latest 
photometric software on the highest quality data. These conditions limit our 
sample to 174 quasars.

\section{Results}
\subsection{Galaxies and quasars}

We study the spatial density of galaxies around quasars at different 
scales and using different limiting magnitudes for the galaxies. As a control 
sample we use the same galaxy catalogs that are used for obtaining the density.

\begin{figure}[h]
  \begin{tabular}{cc}
 \resizebox{\hsize}{!}{\includegraphics{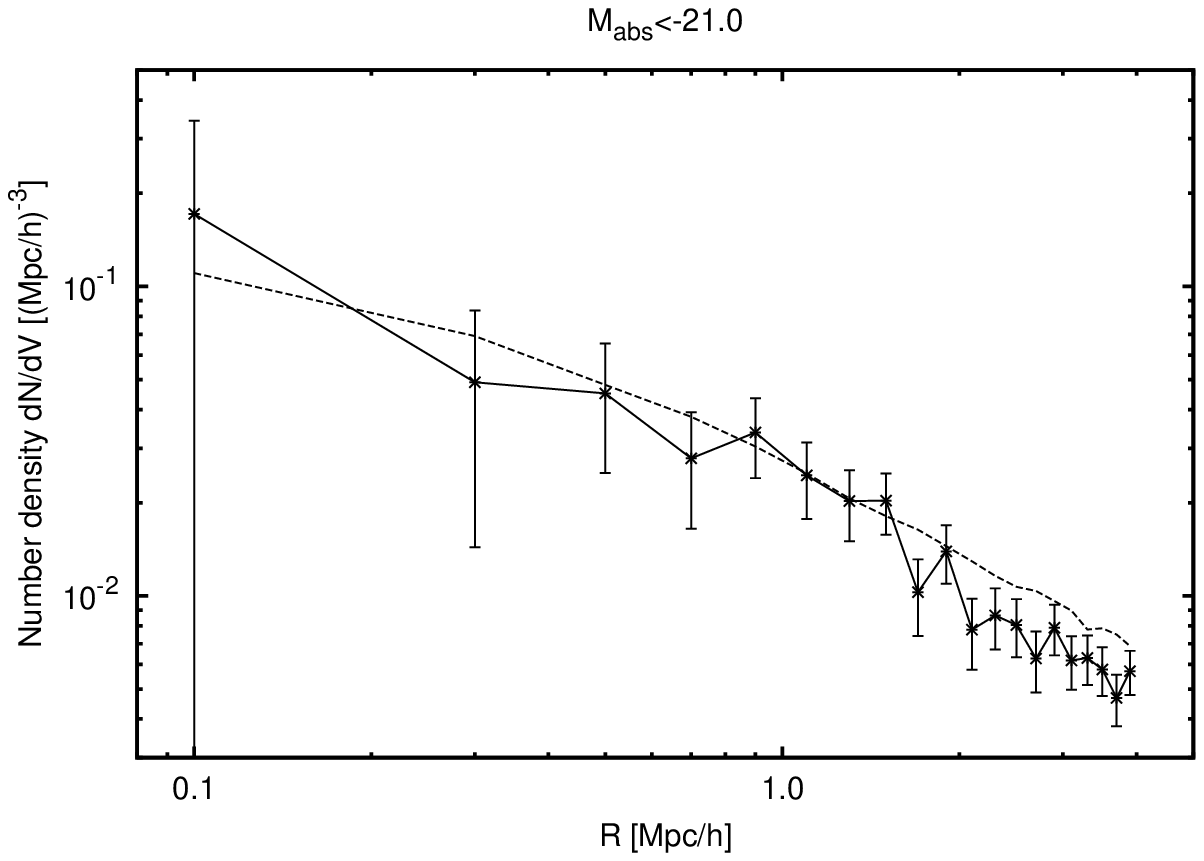}}\\
 \resizebox{\hsize}{!}{\includegraphics{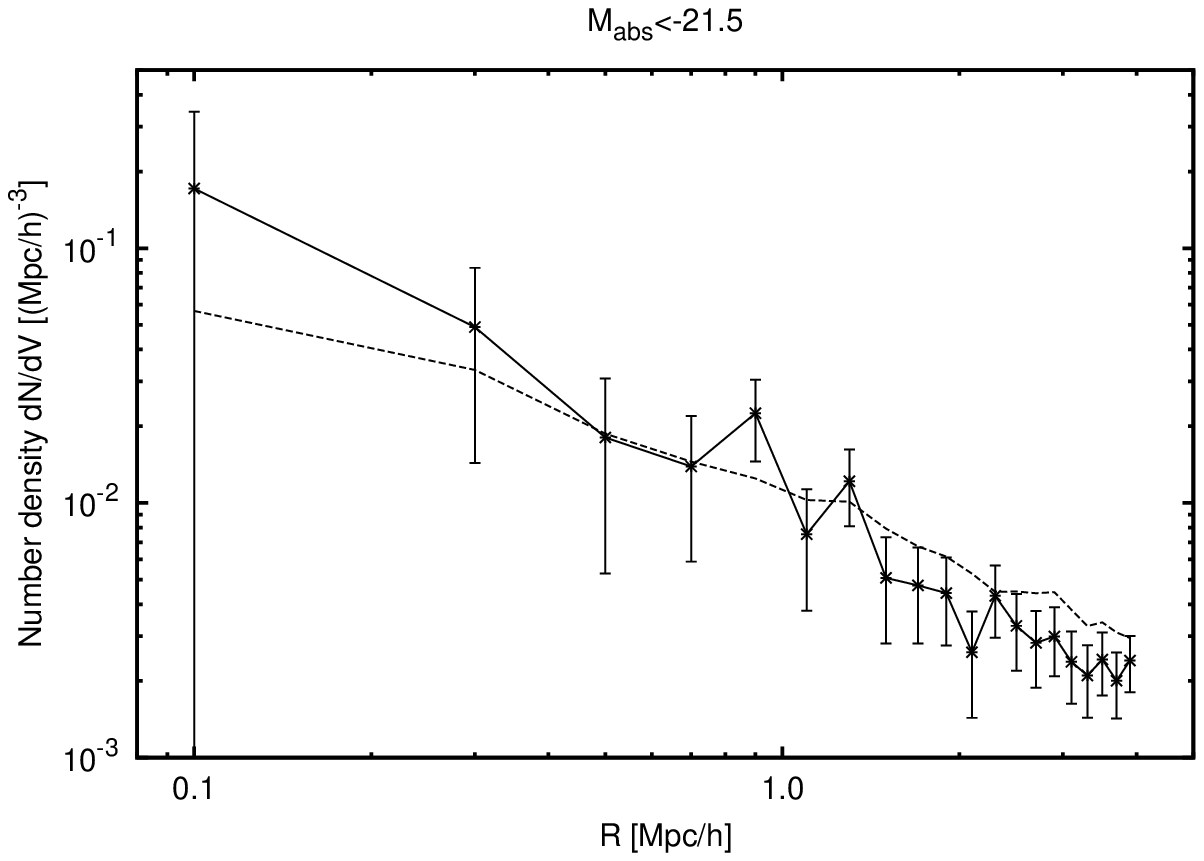}}
  \end{tabular}
  \caption{The spatial density of galaxies at radii $r<4$ \,\Mpc around quasars 
(solid line, with errorbars) and galaxies (dashed line) for the galaxy luminosity 
limits $M<-21.0$ (top) and $M<-21.5$ (bottom). }
  \label{galden4}
\end{figure}

\begin{figure}[h]
  \begin{tabular}{cc}
  \resizebox{\hsize}{!}{\includegraphics{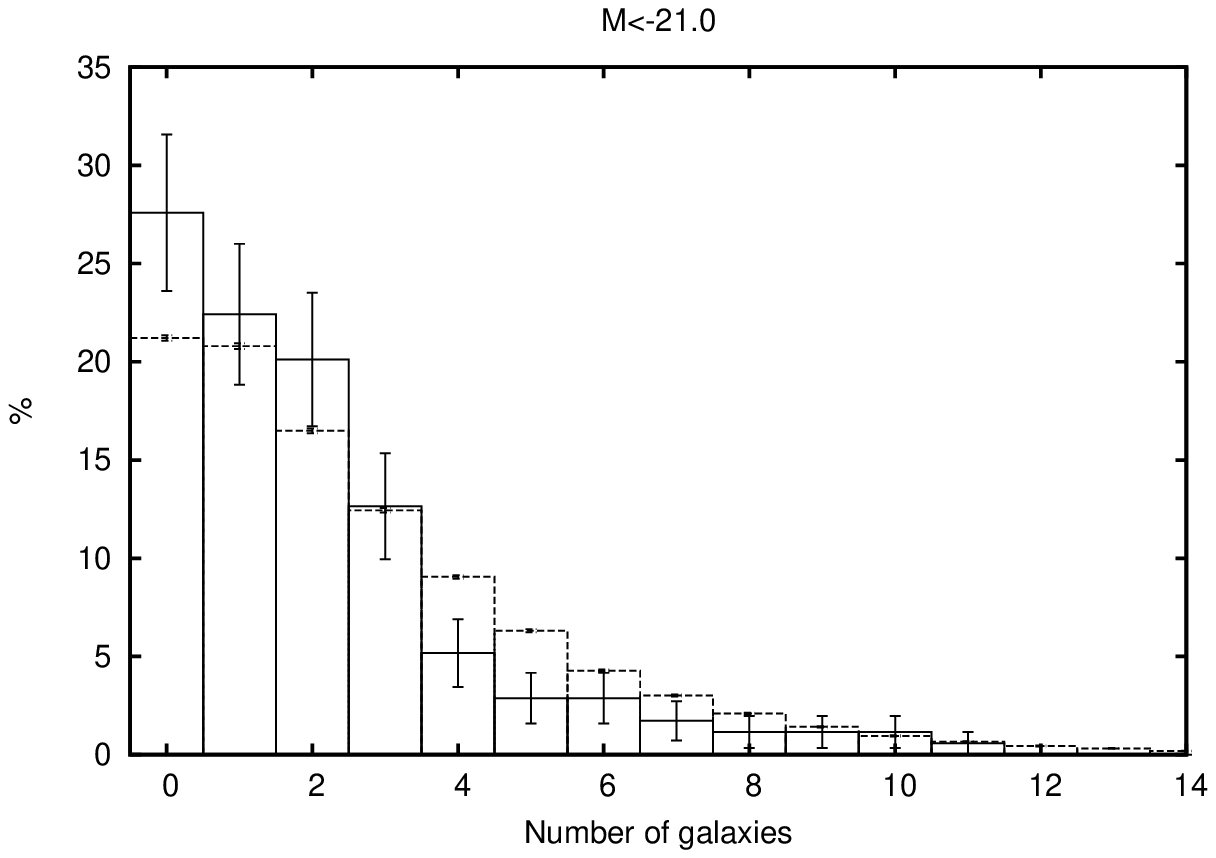}}\\
  \resizebox{\hsize}{!}{\includegraphics{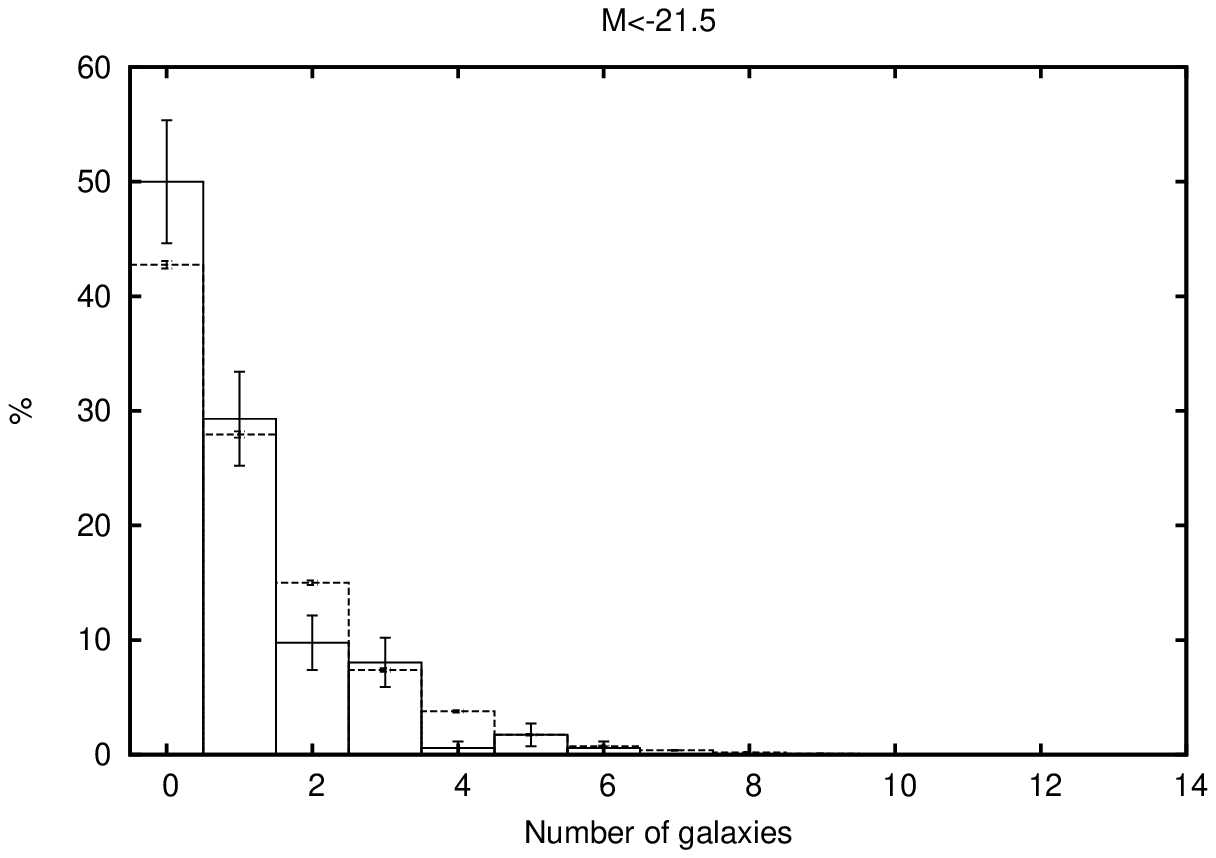}}
  \end{tabular}
  \caption{Distribution of the number of galaxies inside 4\,\Mpc radii from 
quasars (solid line histogram) and galaxies (dashed line histogram) with 
limiting magnitudes $M<-21.0$ (top) and $M<-21.5$ (bottom) for the galaxies.}
  \label{ndist4}
\end{figure}

As a first simple measure of the density we estimate the redshift space
distance from each quasar to the nearest galaxy brighter than 
$M=-21.5$.
The average distance between a quasar and its nearest galaxy is 
$\left(4.9 \pm 0.3\right)$\,\Mpc, while the average 
distance between a $M<-21.5$ galaxy and the nearest other galaxy is 
$\left(4.30 \pm 0.02\right)$\,\Mpc. 
Althoug this estimate is quite uncertain, it may be a hint of 
quasars lying in less dense regions than galaxies 
with $M<-21.5$ in general.

Then we estimate the number density of galaxies at different distances from the 
quasars. We count the number of galaxies in 0.1\,\Mpc bins 
from 0.1 to 4\,\Mpc distance from each of the quasars. 
This number is then divided by the total co-moving volume of this bin for all the 
quasars,
\[
\frac{\mathrm{d}N}{\mathrm{d}V}(R_i)=\frac{\frac{1}{n}\sum_{j=1}^{n}N_j}{V(R_i)-V(R_{i-1})}
\]
where $n$ is the number of quasars, $N_j$ is the number of galaxies at a distance
$R_i$ from each quasar, and $V(R_i)$ is the co-moving volume of a sphere with the radius $R_i$.
This gives the number density of galaxies in this 
distance bin for all the quasars. 
The same calculation is done for the reference galaxies.

Figure~\ref{galden4} shows the mean number density of galaxies as a function 
of distance from quasars and from galaxies. The results hint at a galaxy
overdensity at the radii less than 1\,\Mpc around quasars. 
However, because of the small number of quasars in our sample the Poissonian 
errors at the radii $<4$\,\Mpc are large, and the differences between 
quasars and galaxies are within the error limits.

\begin{figure*}[ht]
\centering
\resizebox{0.8\columnwidth}{!}{\includegraphics*{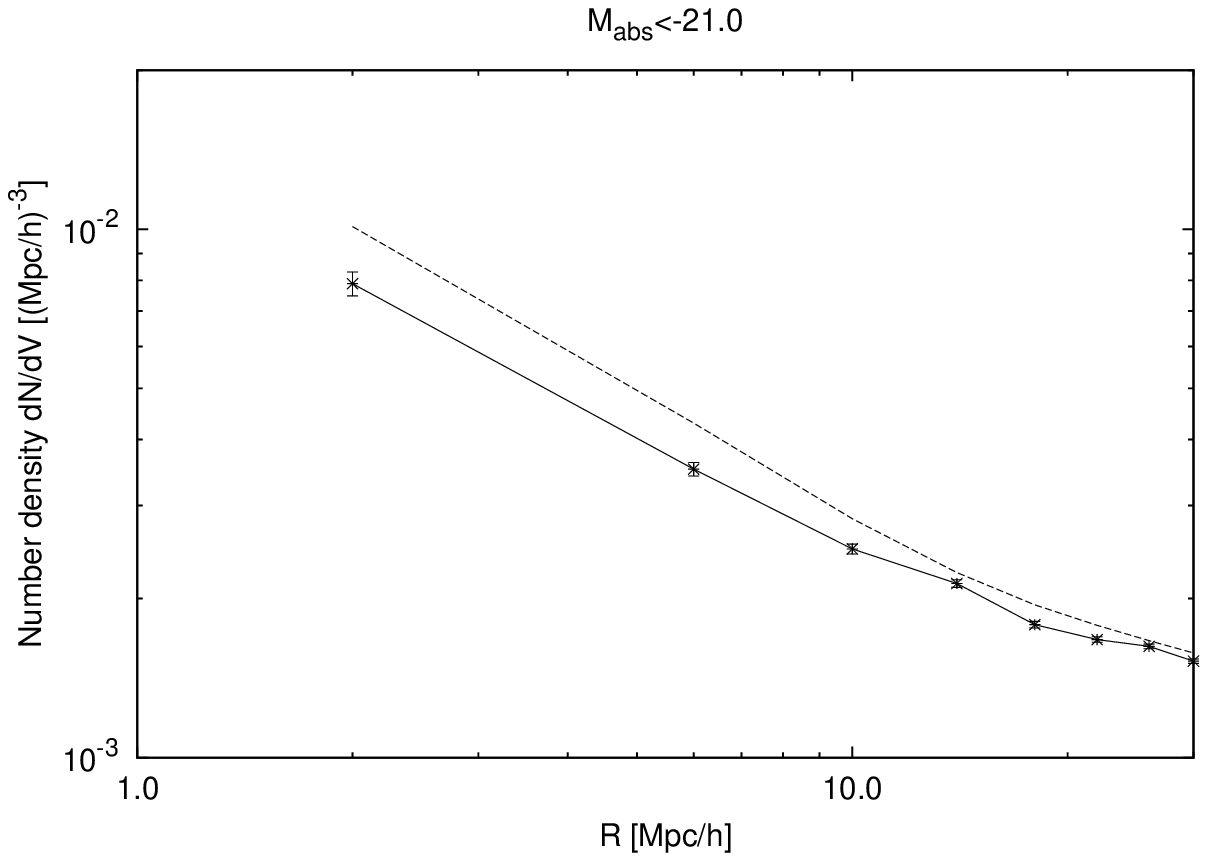}}
\resizebox{0.8\columnwidth}{!}{\includegraphics*{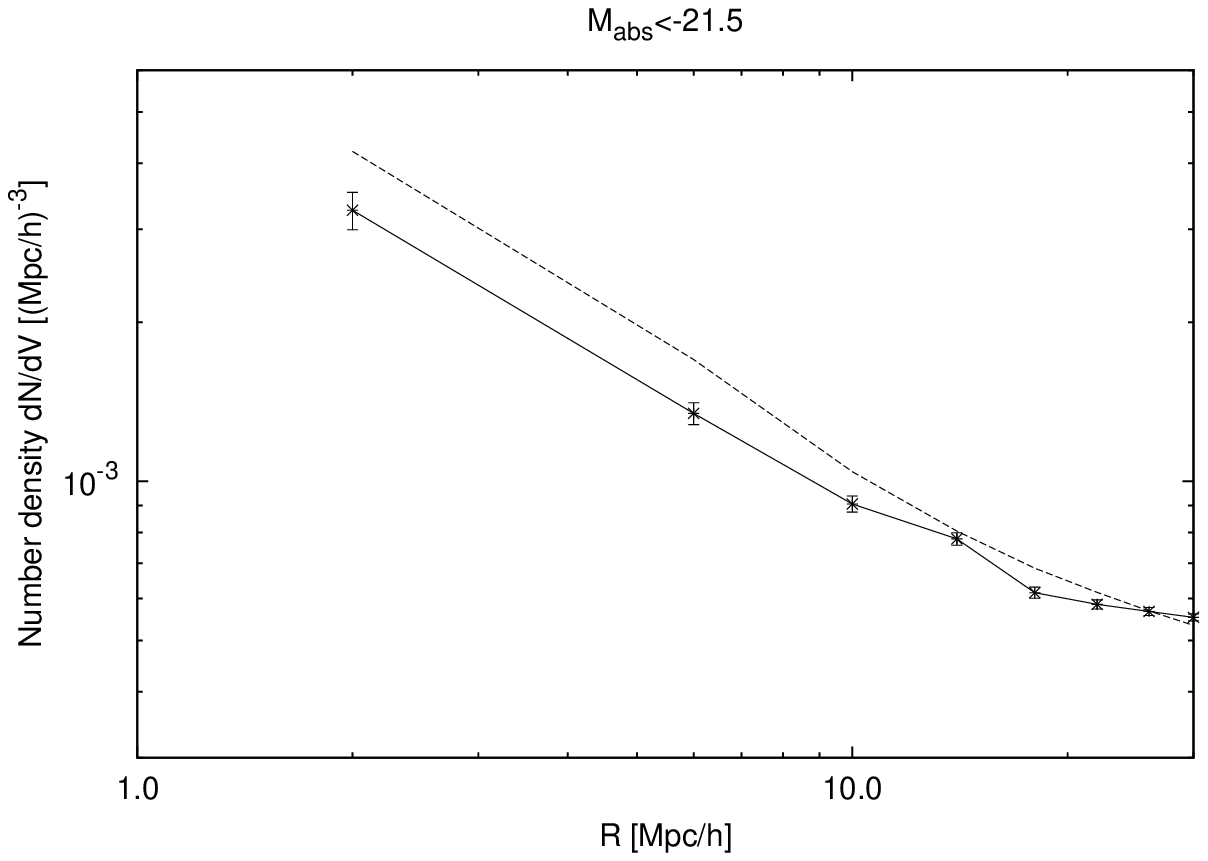}}
\hspace*{2mm}\\
\resizebox{0.8\columnwidth}{!}{\includegraphics*{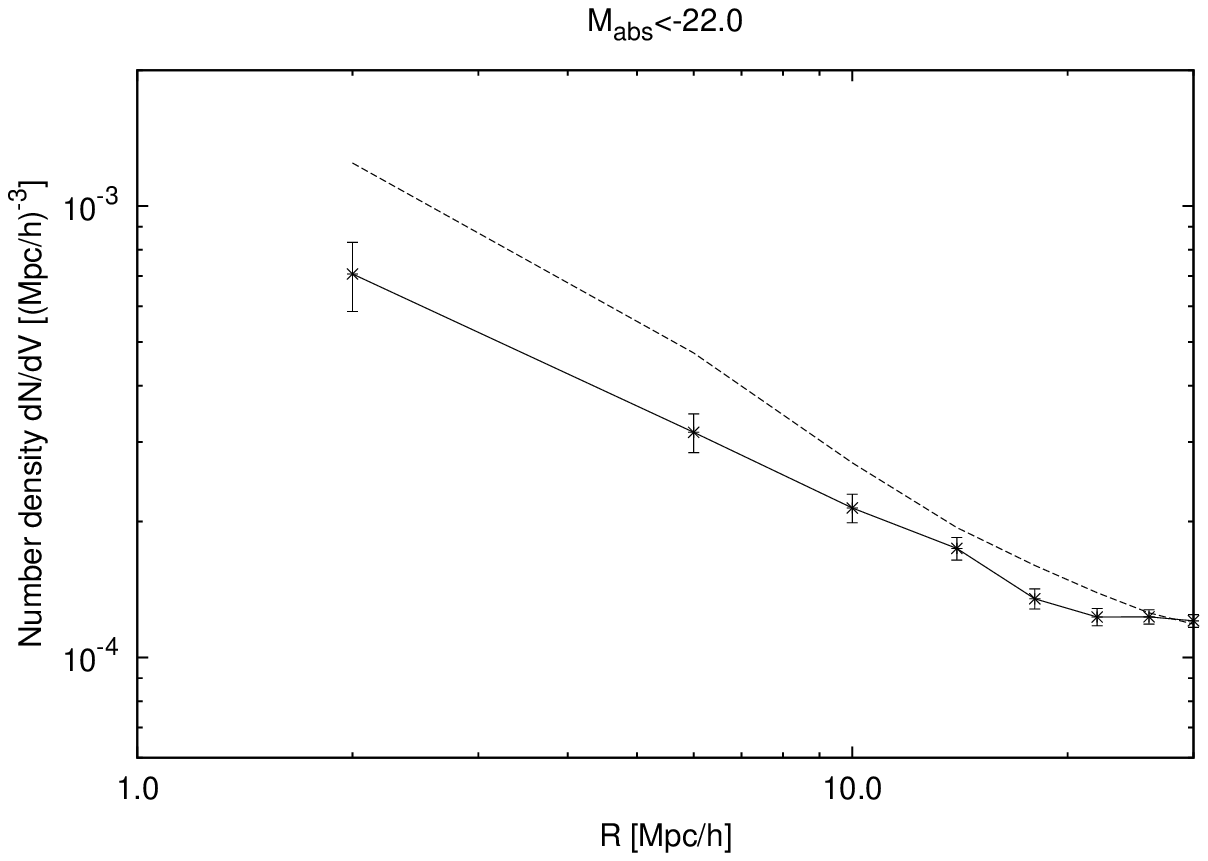}} 
\resizebox{0.8\columnwidth}{!}{\includegraphics*{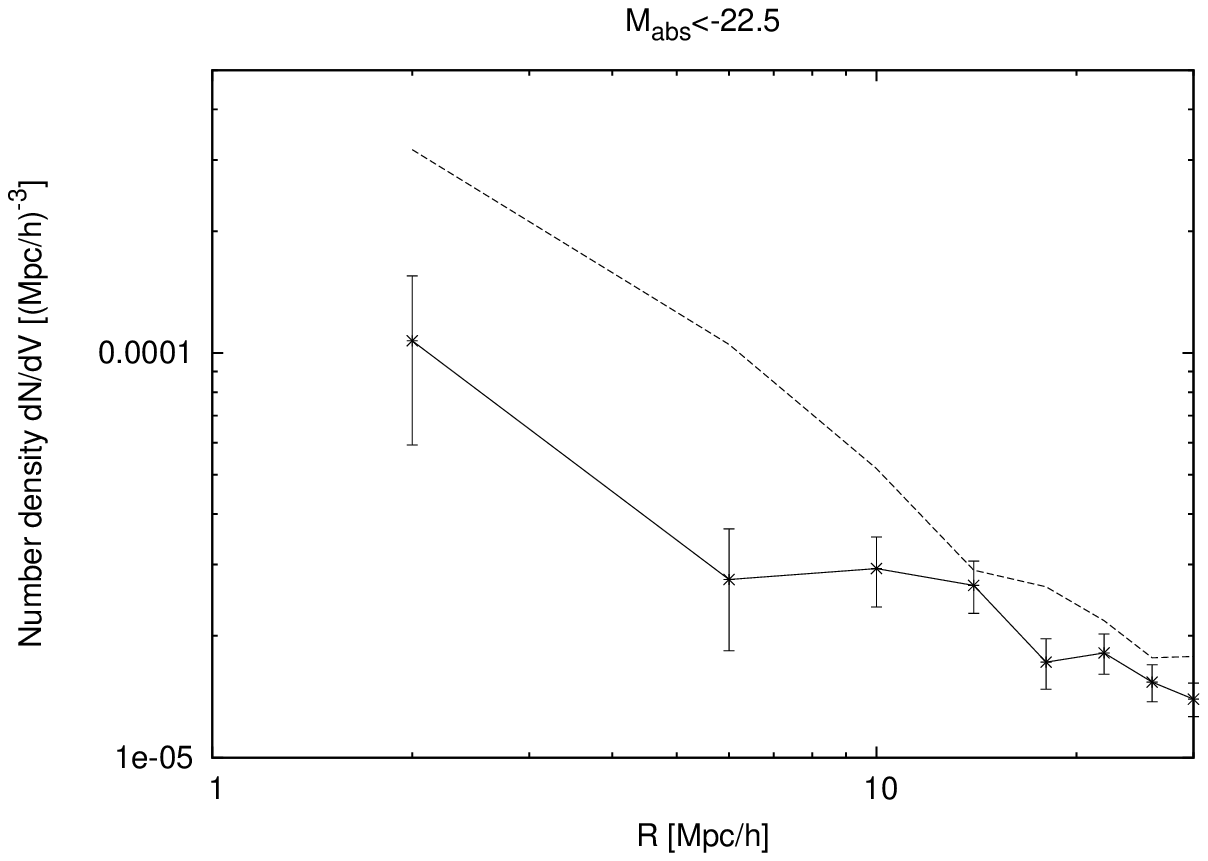}}
  \caption{The spatial density of galaxies at radii $r<60$\,\Mpc around 
quasars (solid line) and galaxies (dashed line) for galaxy luminosity 
limits $M<-21.0$, $M<-21.5$, $M<-22.0$, and $M<-22.5$. }
  \label{galden60}
\centering
\resizebox{0.8\columnwidth}{!}{\includegraphics*{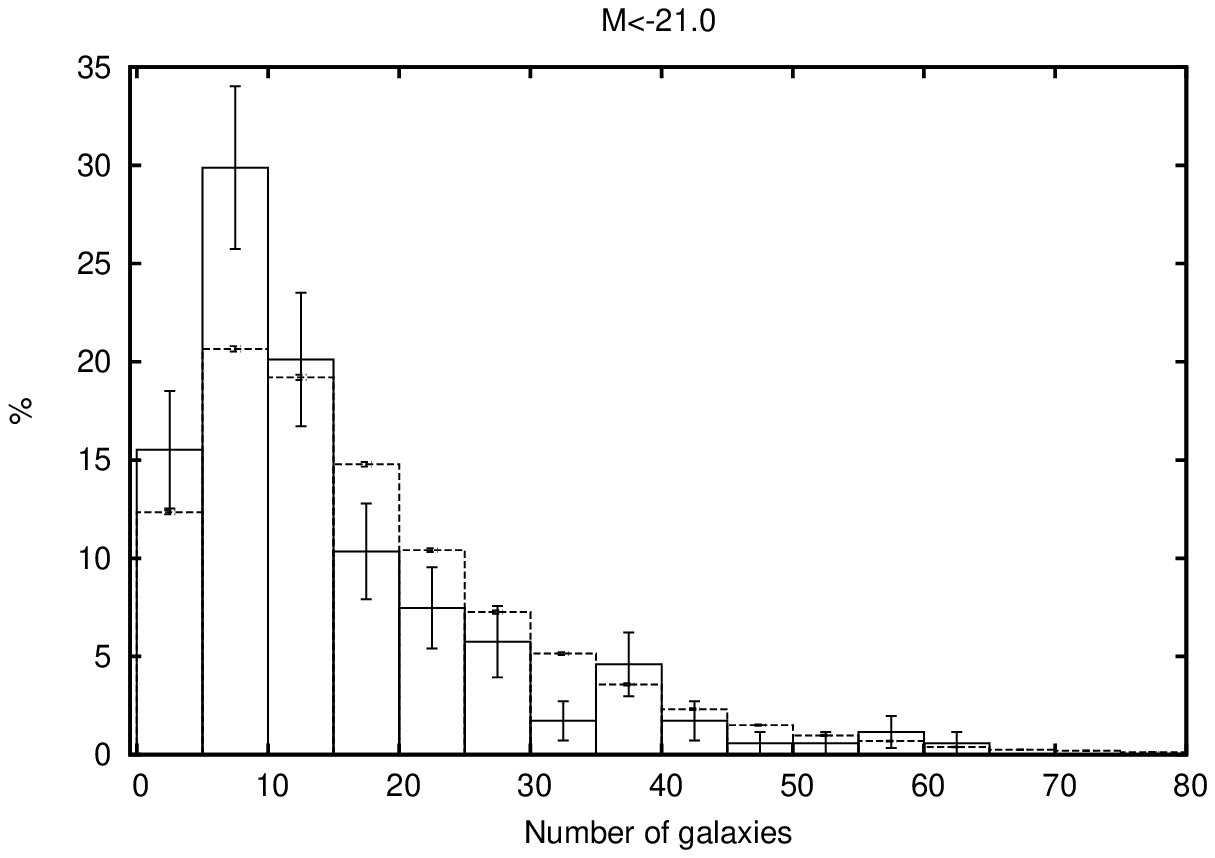}}
\resizebox{0.8\columnwidth}{!}{\includegraphics*{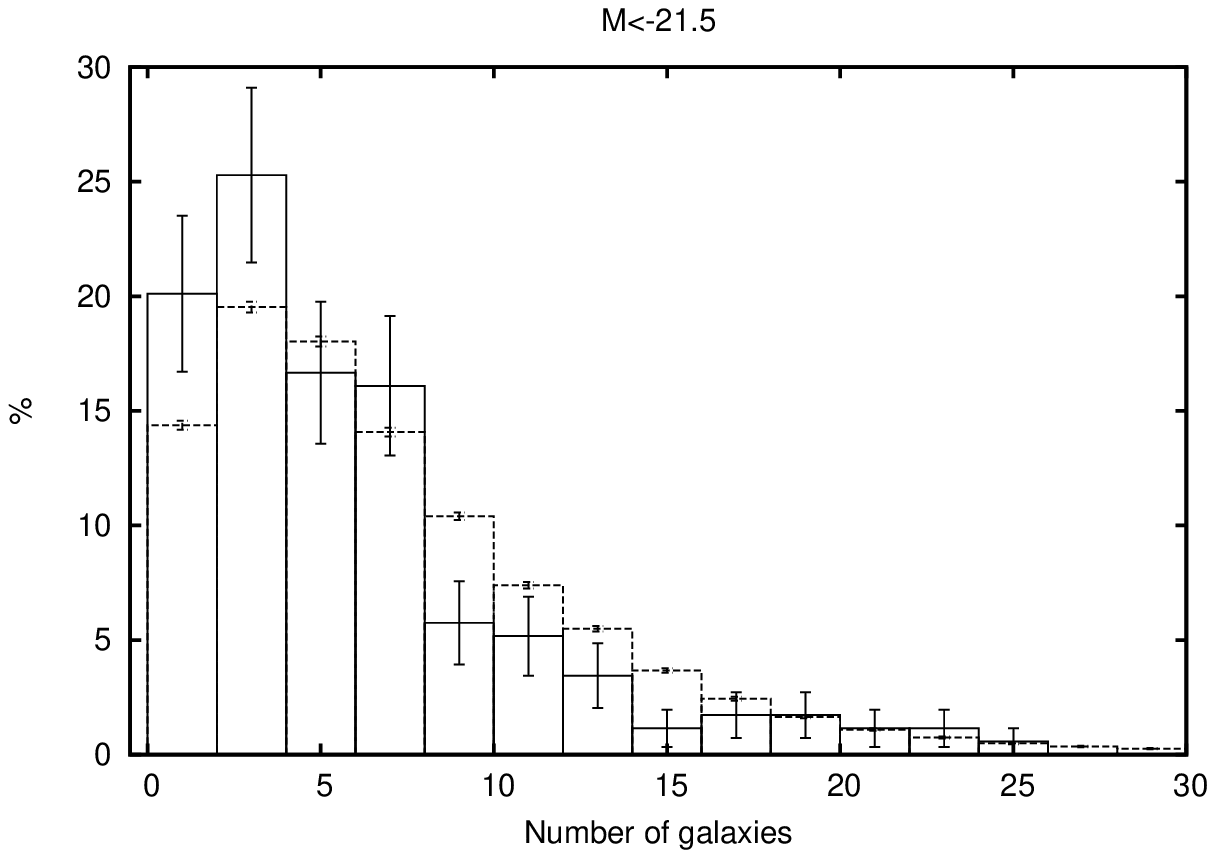}}
\hspace*{2mm}\\
\resizebox{0.8\columnwidth}{!}{\includegraphics*{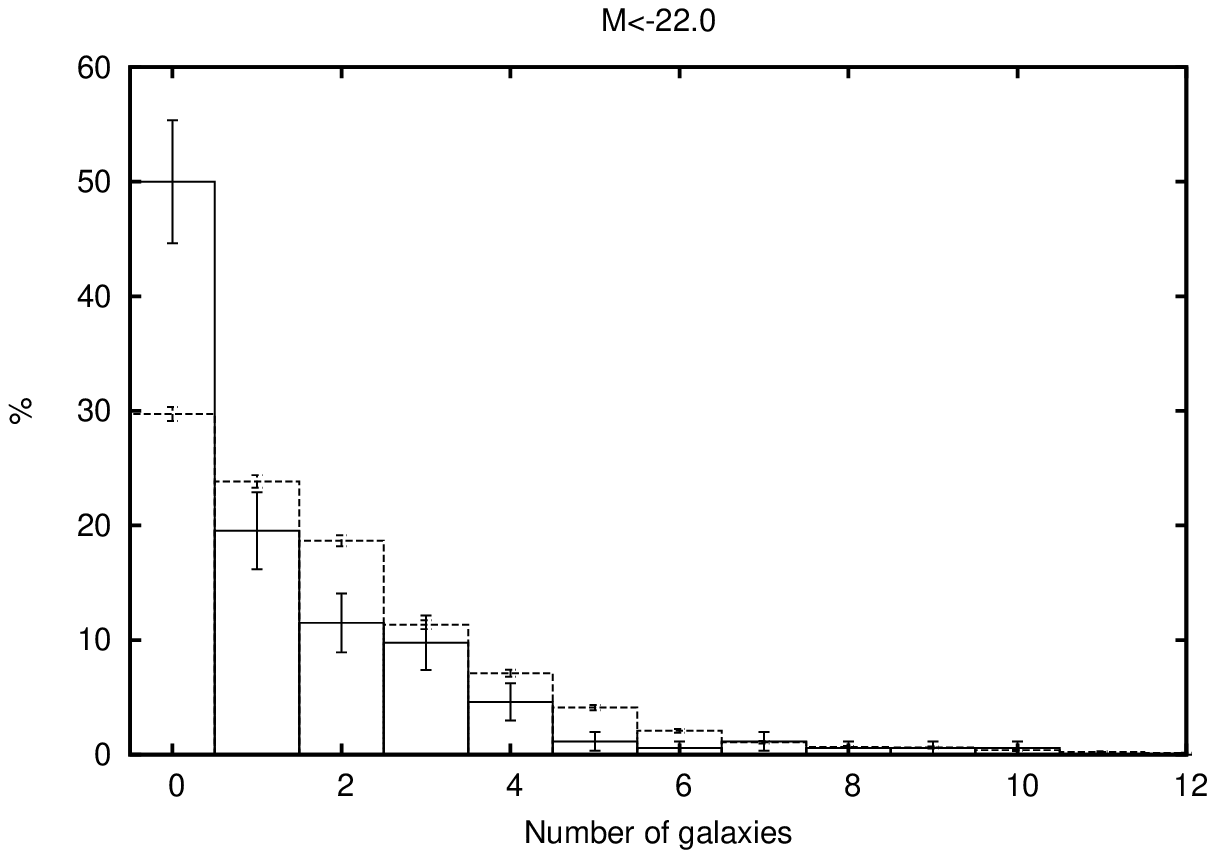}} 
\resizebox{0.8\columnwidth}{!}{\includegraphics*{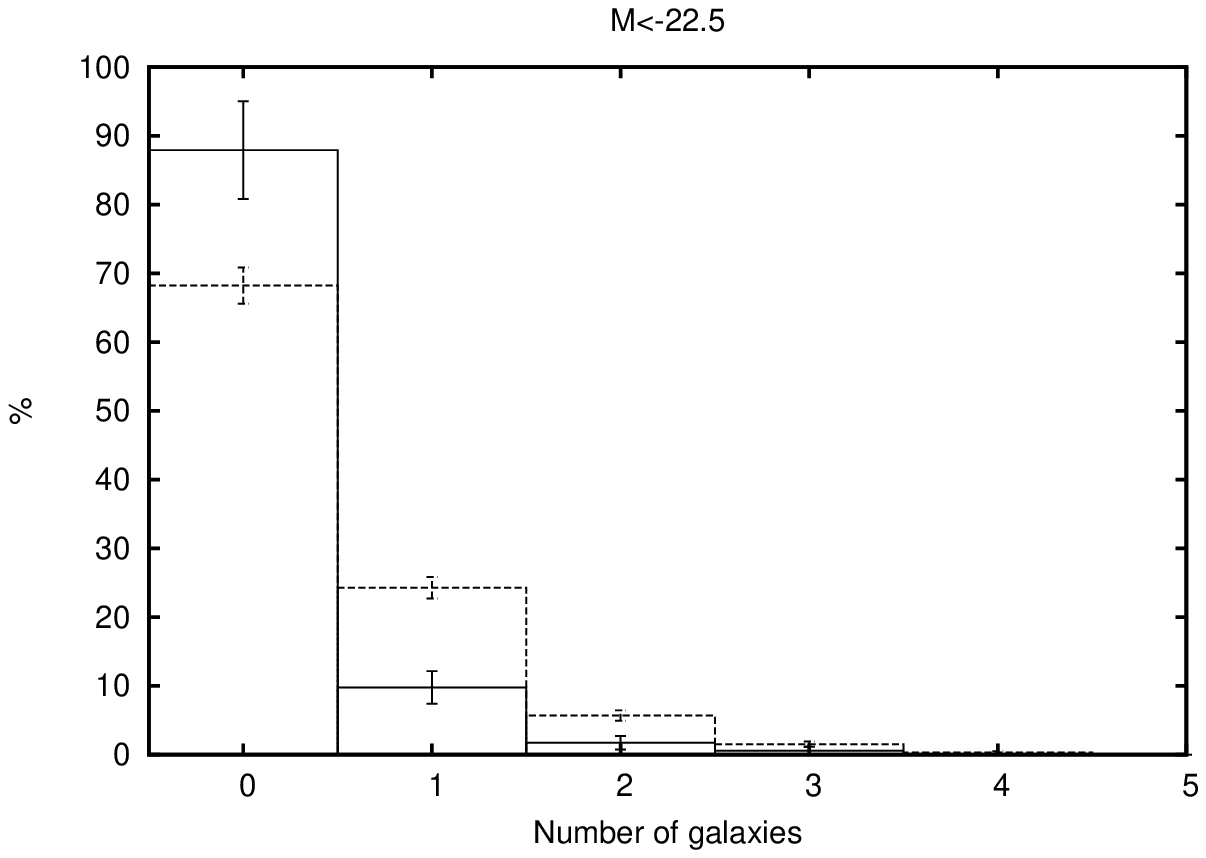}}
\caption{Distribution of the number of galaxies inside 10\,\Mpc radii from 
quasars (solid line histogram) and galaxies (dashed line histogram) with 
limiting magnitudes for galaxies $M<-21.0$ (top left panel),$M<-21.5$ (top right), 
$M<-22.0$ (bottom left), and $M<-22.5$ (bottom right).}
 \label{ndist10}

\end{figure*}

In order to study the galaxy density inside the whole co-moving volume enclosed by the 
4\,\Mpc radius from the quasars we find the distribution of the number of 
galaxies inside this radius.
The results for the environments of quasars and galaxies for the magnitude limits 
$M<-21.0$ and $M<-21.5$ are shown in Fig.~\ref{ndist4}.  
Of the quasars in our sample, 28\,\% have no $M<-21.0$ galaxies inside the 4\,\Mpc 
radius, while for the $M<-21.0$ galaxies themselves, this fraction is only 21\,\%.
The Kolmogorov-Smirnov test (Table~1) shows that the distributions 
are different on the 98.6\,\% significance level.
This suggests that the quasars reside in poorer environments than the 
$M<-21.0$ galaxies. Differences in the distributions for the limiting galaxy
magnitude $M<-21.5$ are less significant, but 
Fig.~\ref{ndist4} shows that the fraction of quasars with no galaxies within the 
4\,\Mpc radius is slightly larger than the fraction of galaxies.  For 
magnitude limits $M<-22.0$ and $M<-22.5$ the number of galaxies is too small to show any differences.

\begin{table}
\centering
\begin{tabular}{c c c c}
\hline\hline
Magnitude Limit & Distance Limit [\Mpc] & $\hat{D}$ &Probability \\
\hline
$-21.0$ & 4 & 0.118 & 0.01442\\
$-21.5$ & 4 & 0.086 & 0.1467\\
$-21.0$ & 10& 0.141 & $1.748\times 10^{-3}$\\
$-21.5$ & 10& 0.139 & $2.176\times 10^{-3}$\\
$-22.0$ & 10& 0.203 & $1.227\times 10^{-6}$\\
$-22.5$ & 10& 0.197 & $1.621\times 10^{-5}$\\
\hline
\end{tabular}
\label{KStest}
\caption{Kolmogorov-Smirnov test statistics for distributions of the number of 
galaxies inside 4 and 10\,\Mpc radii around quasars and galaxies. Probabilities
for the distributions to be similar are shown in the fourth column. }
\end{table}

Figure~\ref{galden60} shows the spatial density of galaxies around quasars 
on larger scales, from 2\,\Mpc to 30\,\Mpc. This was calculated in the same way 
as the galaxy density on smaller scales, but the bin is now 4\,$h^{-1}$Mpc.  
Here we use two more magnitude limits 
for galaxies, $M<-22.0$ and $M<-22.5$.
On this scale quasars are 
located in less dense regions than galaxies on average. When the galaxy 
luminosity limit increases from $M<-21.0$ to $M<-22.5$, the difference between the
quasar and galaxy environments grows.


We also calculated the number densities for a volume-limited sample of galaxies with $M<-20.0$.
This sample extends only up to 316\,\Mpc which limits the number of quasars to 
11. Underdensity around quasars can be detected, but there are not enough data 
to give accurate results.

The underdensity around quasars can also be seen in Fig.~\ref{ndist10} that
shows that the number of galaxies closer than 10\,\Mpc to quasars tends to be 
lower than the number of other galaxies closer than 10\,\Mpc to galaxies.
For $M<-21.0$ galaxies, 45\,\% of quasars have less than ten galaxies 
within the 10\,\Mpc distance, while of the $M<-21.0$ galaxies themselves, only 
33\,\% belong to this group. The effect is stronger for brighter galaxies.
For the magnitude limit $M<-22.0$, 50\,\% of quasars have no galaxies within the 
10\,\Mpc distance, while 30\,\% of the $M<-22.0$ galaxies have no other galaxies 
with the same magnitude limit closer than 10\,\Mpc. As we can see in 
Table~1, the distributions at all the magnitude limits are different 
at very high levels of significance.

\begin{figure}[ht]
\centering
\resizebox{1.0\columnwidth}{!}{\includegraphics{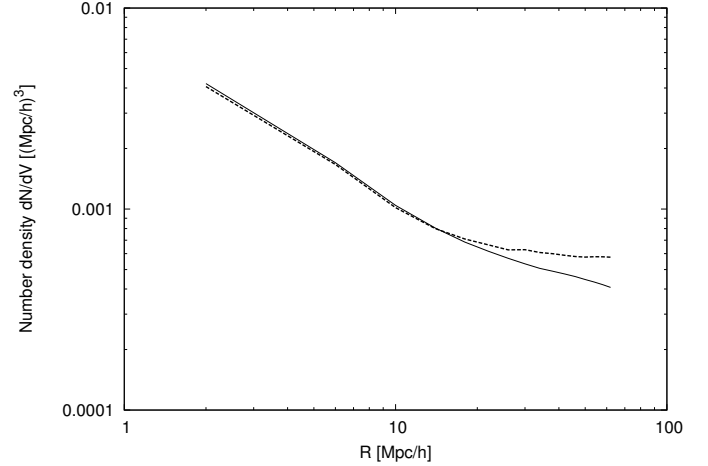}}
  \caption{Edge effects in the galaxy samples. Number density of all galaxies 
  is shown with the solid line, and number density of galaxies in the 15 
  subsamples with the minimum distance to edges of the whole galaxy sample 
  of 60 \Mpc or more with the dashed line.}
\label{edges}
\end{figure}

We have made the number density analysis also for a smaller subsample 
of the galaxy catalog in order to estimate the effect of the edges of the 
sample volume. We chose 15 subsamples from the galaxy catalog with the limiting 
magnitude $M<-21.5$, such that all the galaxies in the subsamples lie at 
least at 60\,\Mpc from the edges of the full sample. The number densities 
of these galaxies compared to the whole galaxy sample is shown in Fig. 
\ref{edges}.
This analysis showed us that the estimate of the
number density is not significantly biased by the edges of the sample for 
distances less than 20\,\Mpc. For the neighbor distance 50\,\Mpc our estimate is approximately 30\,\% 
too small.

\begin{figure*}[ht]
\centering
\resizebox{0.75\columnwidth}{!}{\includegraphics*{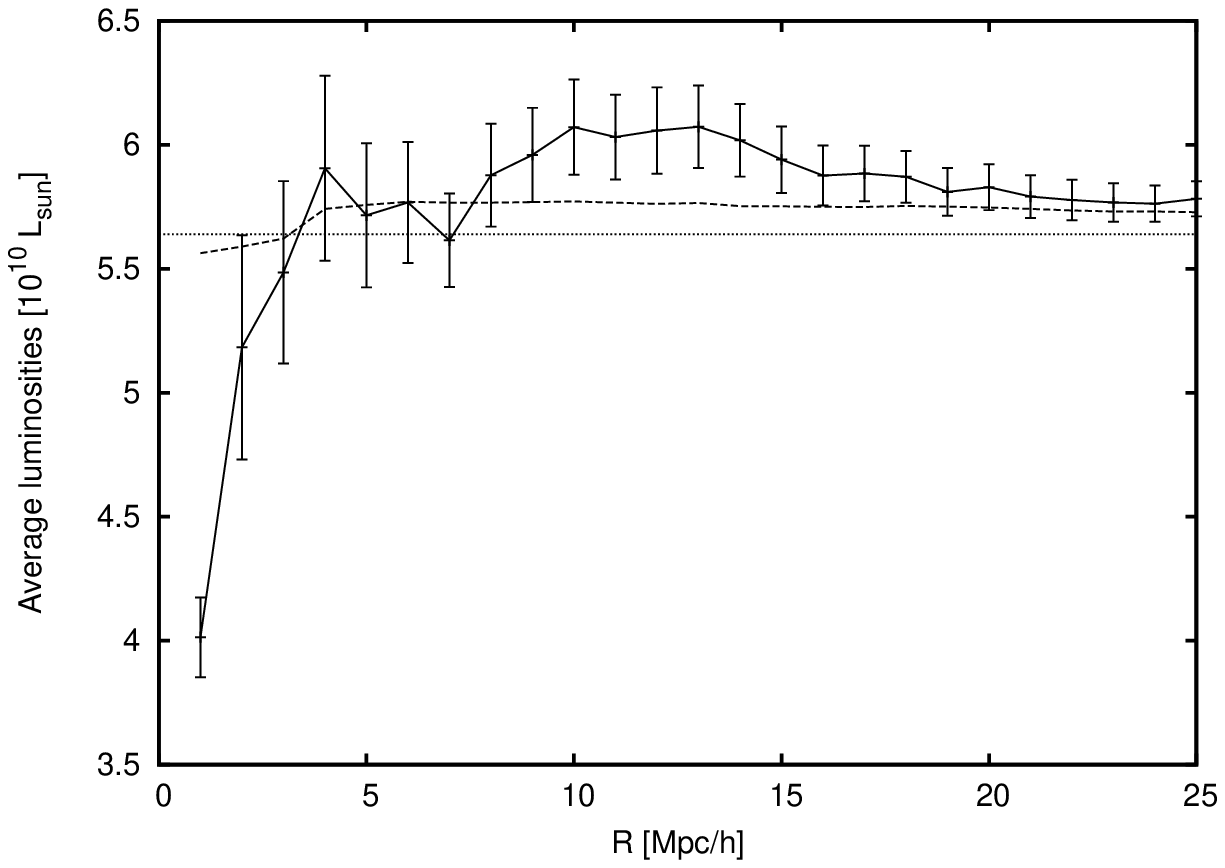}}
\resizebox{0.75\columnwidth}{!}{\includegraphics*{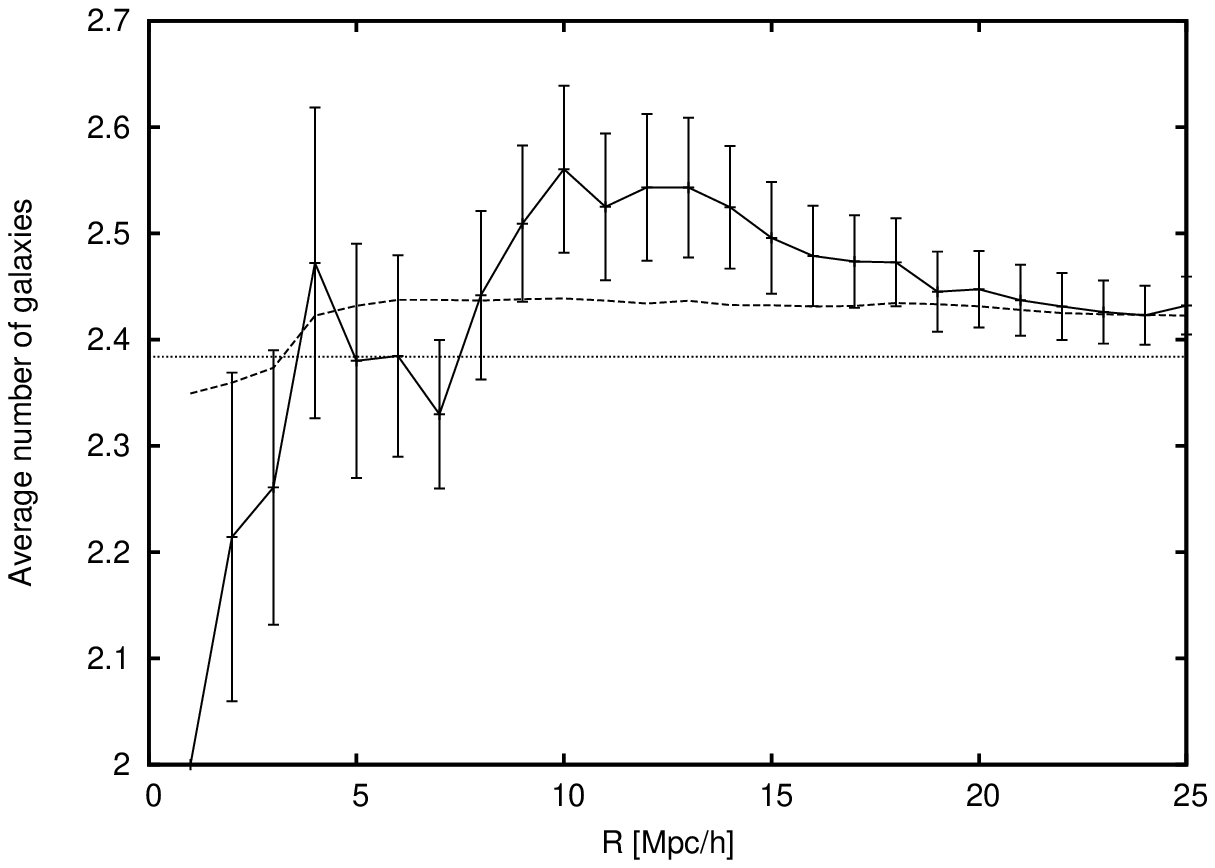}}
\hspace*{2mm}\\
\resizebox{0.75\columnwidth}{!}{\includegraphics*{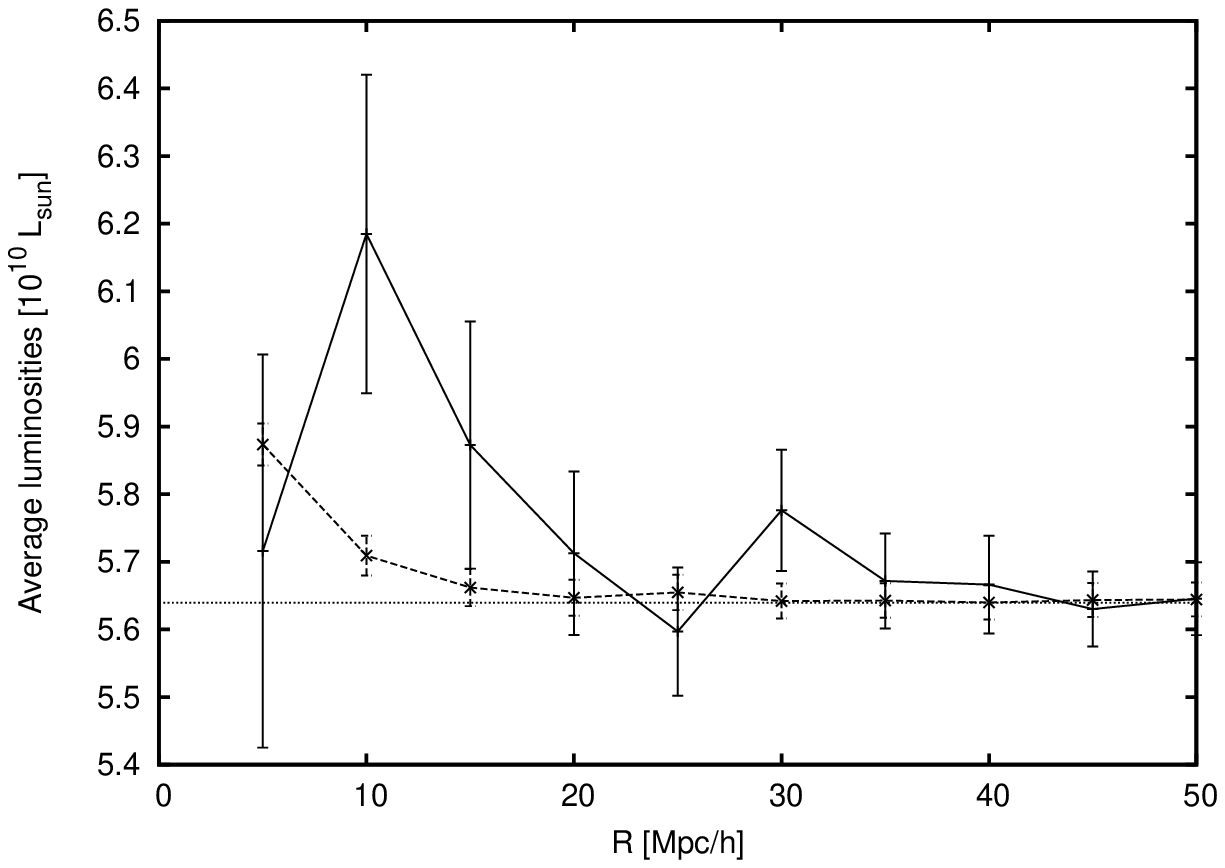}}
\resizebox{0.75\columnwidth}{!}{\includegraphics*{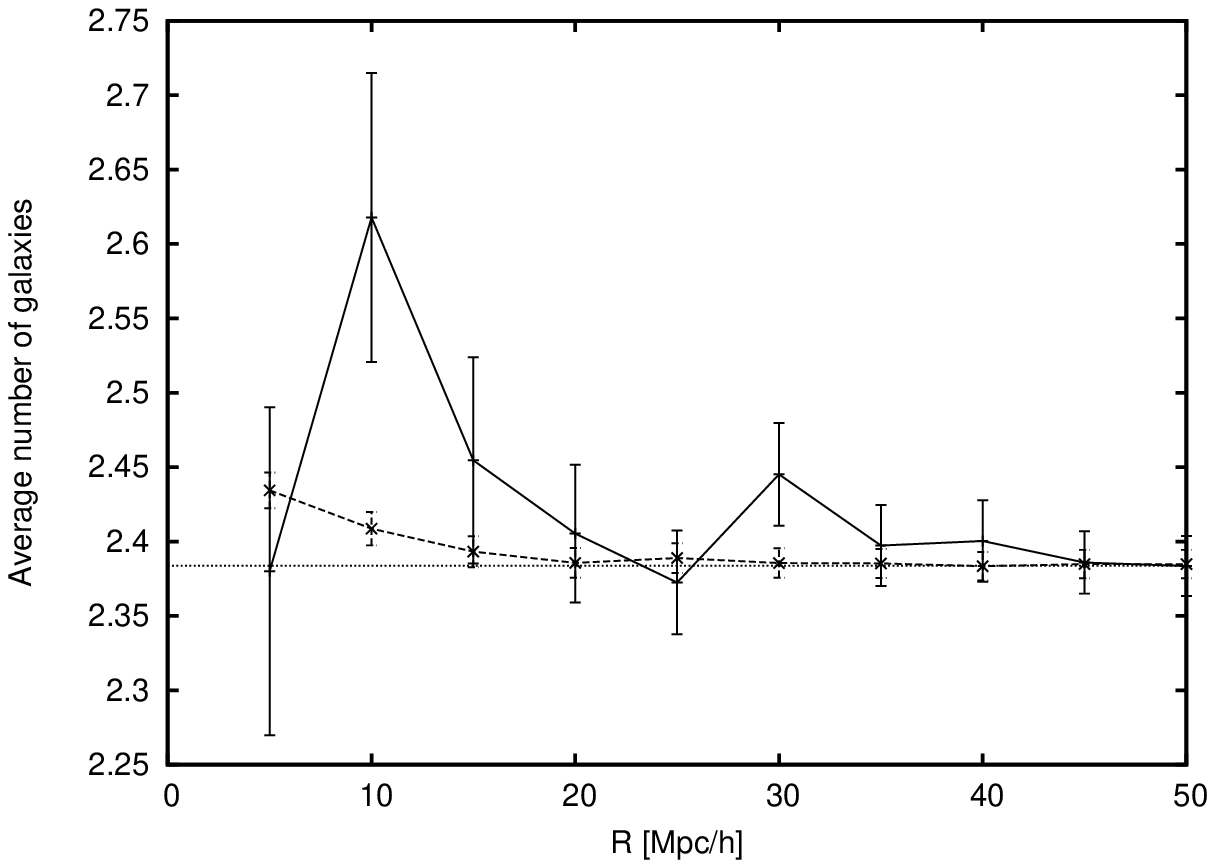}}
  \caption{Bottom: Average luminosities (left) and numbers of galaxies (right) of 
groups that have quasar neighbors (solid line) or galaxy neighbors (dashed line) between the 
radii $R-5$\,\Mpc and $R$. 
The average luminosity and number of galaxies of all groups in the 
sample are shown with dotted lines.
Top: Average luminosities (left) and numbers of galaxies (right) of 
groups that have quasar neighbors (solid line) or galaxy neighbors (dashed line) inside 
the radius $R$. The average luminosity and number of galaxies of all groups in the 
sample are shown with dotted lines.}
  \label{grlumnumdi}
\centering
\resizebox{0.75\columnwidth}{!}{\includegraphics*{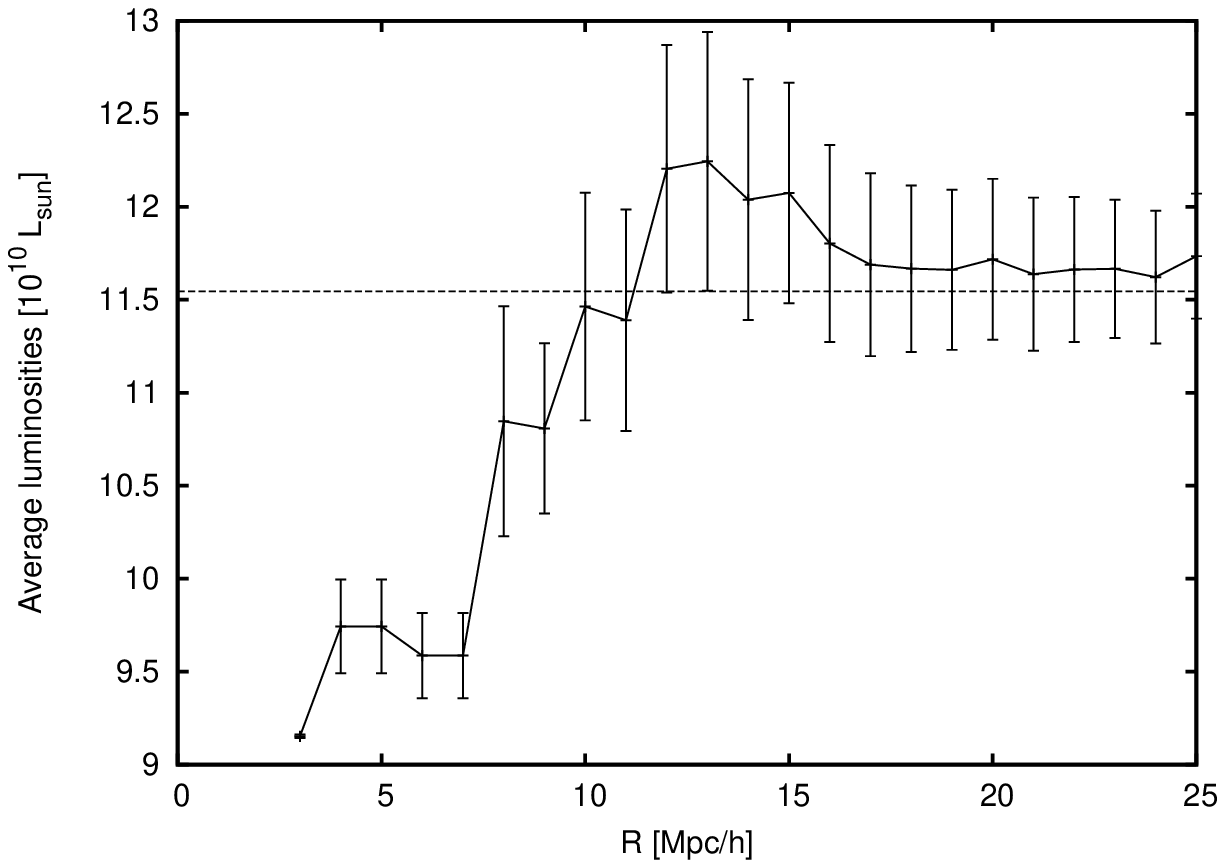}}
\resizebox{0.75\columnwidth}{!}{\includegraphics*{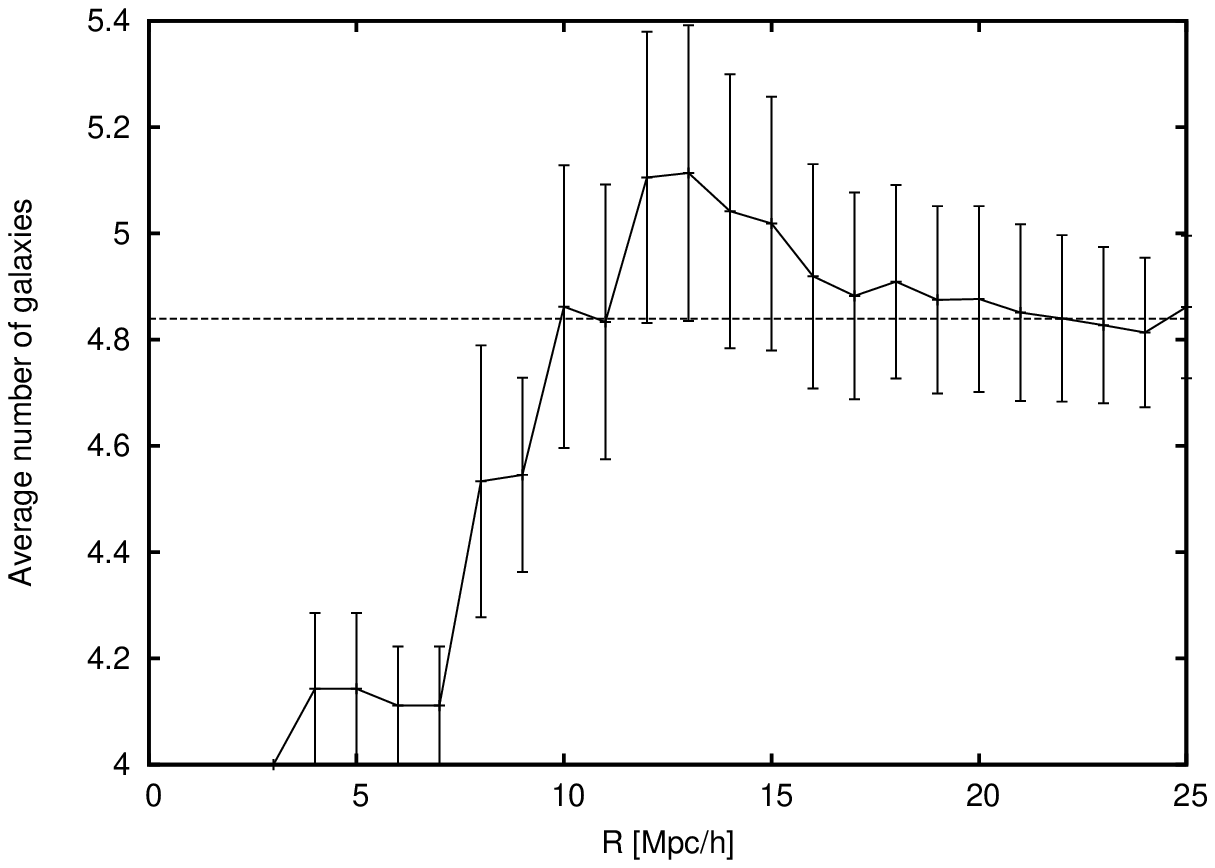}}
\hspace*{2mm}\\
\resizebox{0.75\columnwidth}{!}{\includegraphics*{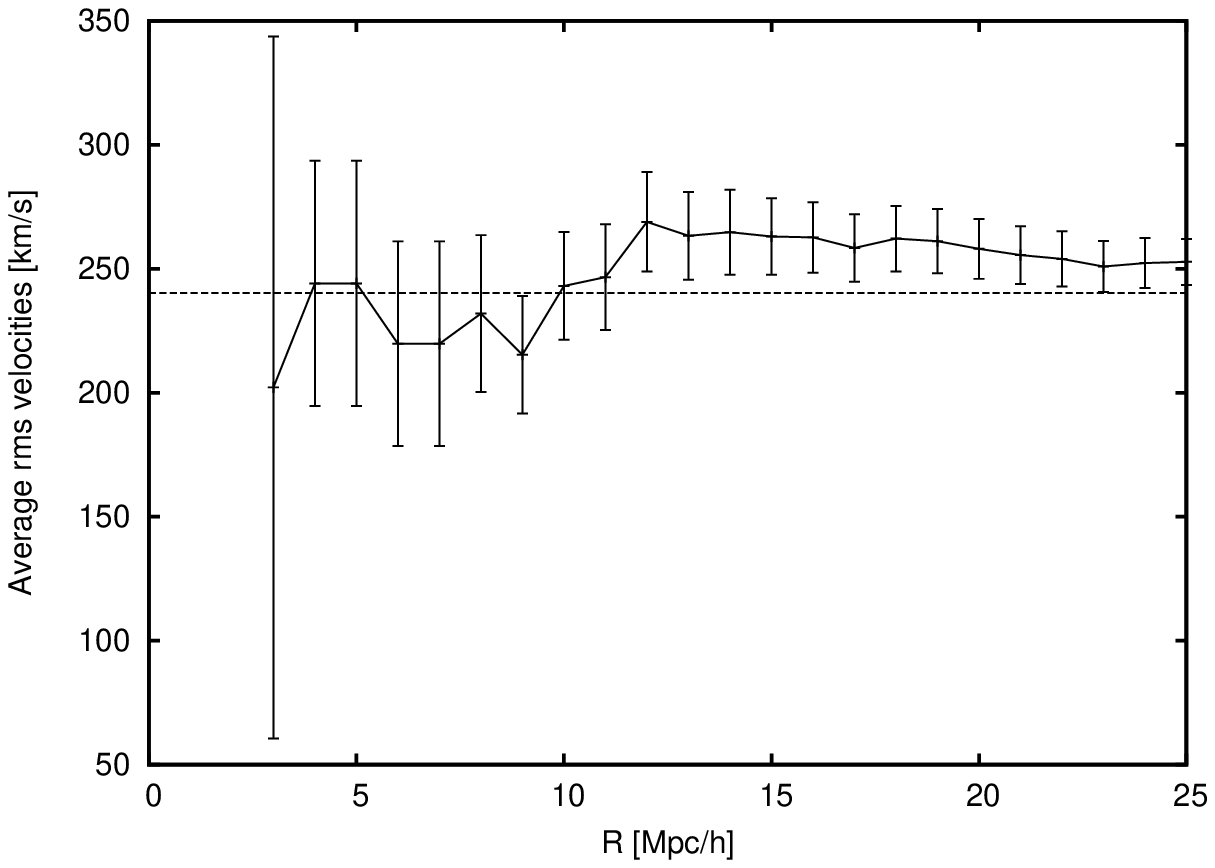}}
  \caption{Average luminosities (top left), numbers of galaxies (top right),
 and rms velocities (bottom) of groups 
of at least 4 galaxies that have quasar neighbors inside the radius $R$. The averages 
for all rich groups in the sample are shown with dashed lines.
}
  \label{grrich}
\end{figure*}

\subsection{Groups and quasars}

Our volume limited group catalog contains 9581 groups, but a great majority 
of them are groups of only two galaxies. In some cases we will
use a subsample of the 777 groups that have 4 or more galaxies.
This way we can see if the properties of rich groups near quasars 
differ from the properties of poor groups.

We study the distribution of groups of galaxies in the environments of 
quasars in a similar manner as we did for the distribution of galaxies.
The average distance from a quasar to the nearest group is $(7.8 \pm 0.5)$\,\Mpc 
and the average distance from a galaxy with $M<-21.5$ to the nearest group is 
$(6.49 \pm 0.04)$\,\Mpc. This result suggests that quasars are farther from 
the groups than galaxies on average. However, if we compare only the distances 
to the richer groups we find no difference at all: The average distance from a
quasar to the nearest group of at least 4 galaxies is $(22 \pm 2)$\,\Mpc, while 
the average distance from a $M<-21.5$ galaxy to the nearest rich group is 
$(21.64 \pm 0.08)$\,\Mpc.

The average luminosities and numbers of galaxies of groups that have 
quasar or galaxy neighbors at a given distance are shown in Fig.~\ref{grlumnumdi}. 
The galaxy sample used for comparison has the limiting magnitude $M<-21.0$.
The plots in the top part of Fig.~\ref{grlumnumdi} are cumulative; e.g., the values 
at $R=5$\,\Mpc are the averages for all the groups that have at least one quasar 
or galaxy neigbor closer than 5\,\Mpc. The mean luminosity of all the 
groups in the sample is $5.64 \times 10^{10} L_{\odot}$, and the mean number 
of galaxies in the groups is 2.38. Both curves approach the average for all 
groups as the neighbor distance grows.

Figure~\ref{grlumnumdi} suggests that the groups that have a quasar at a distance 
less than 2\,\Mpc have very low luminosities and small numbers of galaxies.
This result is uncertain because the number of these groups is small, only 14 
groups have quasars at $R<2$\,$h^{-1}$Mpc. Another feature that can be seen in the 
figure is that the groups that have quasar neighbors at distances of 10 to 15\,\Mpc 
are brighter and richer than the groups that have bright galaxy neighbors at the
same distances. This can be seen also in the differential plot, shown in the 
bottom panels of Fig.~\ref{grlumnumdi}: the richest and most luminous groups 
are the ones that 
have a quasar neighbor at distances from 5 to 15\,$h^{-1}$Mpc. However, the differences are 
small, and because of the small number of quasars, it is uncertain if the 
correlations are real.

In groups of only two or three galaxies the velocity dispersion 
does not tell much about the 
actual physics of the group. If we consider the subsample of rich groups of 
four or more galaxies, we can study dynamical properties of the groups.
Figure~\ref{grrich} shows the average luminosities, numbers of galaxies, 
and rms velocities for the rich groups that have quasar neighbors closer than 
$R$.




If we consider only the rich groups, we see the same rise in the luminosity 
and the number of galaxies as we saw in the case of all the groups that have 
quasar neighbors closer than 10\,\Mpc. The rich groups that are very close to a
quasar are clearly poorer and less luminous than the rich groups in general. 
The richest and the most luminous groups  
have their closest quasar at approximately 10--15\,\Mpc from the group. 
This effect can also be seen in the rms velocities, 
although it is not as clear as in the luminosity and the number of 
galaxies. 

The rise in the group richness and luminosity at the 10--15\,\Mpc distance from quasars 
could be explained by the distribution of quasars within the large scale structure. 
The richest groups usually lie in the cores of 
superclusters, and 10\,\Mpc is a typical distance between a core and a filament 
(Einasto \cite{EinastoM07}). Therefore, this result could be explained by
quasars being located preferentially in filaments.

\subsection{Superclusters and quasars}

The largest density enhancements in the Universe,
superclusters and their environments, 
can be studied using the luminosity density field of galaxies 
(Einasto et al. \cite {Einasto03}). 
It has been shown that the morphological properties of galaxies depend on their 
large-scale cosmological environment (Einasto et al. \cite{EinastoJ07} and 
references therein).  
Both the local (group or cluster scale) 
and global (supercluster scale) environments influence galaxy morphology and 
their star formation activity. Rich superclusters contain a higher 
fraction of early type, passive, red galaxies than poor superclusters
(Einasto et al. \cite{EinastoM07}).  
Inside rich superclusters galaxy properties also vary. The core regions of 
rich superclusters contain a larger fraction of early-type red galaxies and 
richer groups than the outskirts of superclusters (Einasto et al. \cite{EinastoM08}). 
Galaxy 
clusters in the high-density environments are about 5 times more luminous than 
in the low-density environments (Einasto et al. \cite{EinastoJ05}).

The luminosity density field smoothed to appropriate scales
allows us to study the locations of nearby quasars in the supercluster-void network.
Voids, filaments and superclusters 
have characteristic density levels 
within the overall large-scale structure; thus, the density level
can be used to discriminate between different details of the network.
In this procedure we use all the SDSS DR5 galaxies from the Tago
catalog of groups and isolated galaxies (Tago et al. \cite {Tago}).
This catalog is suitable for tracing the supercluster-void network; 
groups are mostly density enhancements within filaments and rich 
clusters are high density peaks of the galaxy distribution in 
superclusters (Tago et al. \cite {Tago} and references therein).  

\begin{figure*}[ht]
\centering
\resizebox{0.9\textwidth}{!}{\includegraphics*{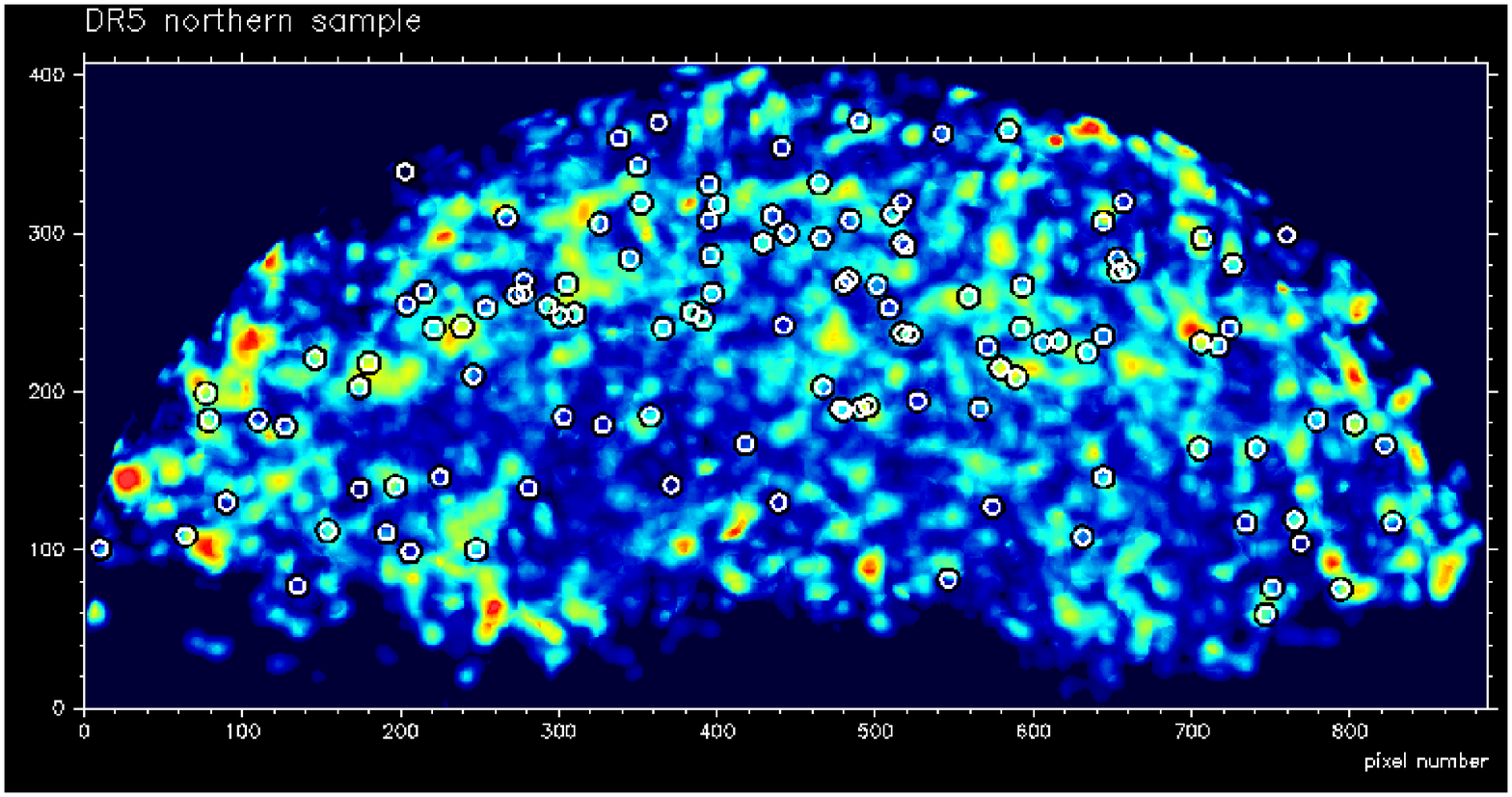}}
\resizebox{0.9\textwidth}{!}{\includegraphics*{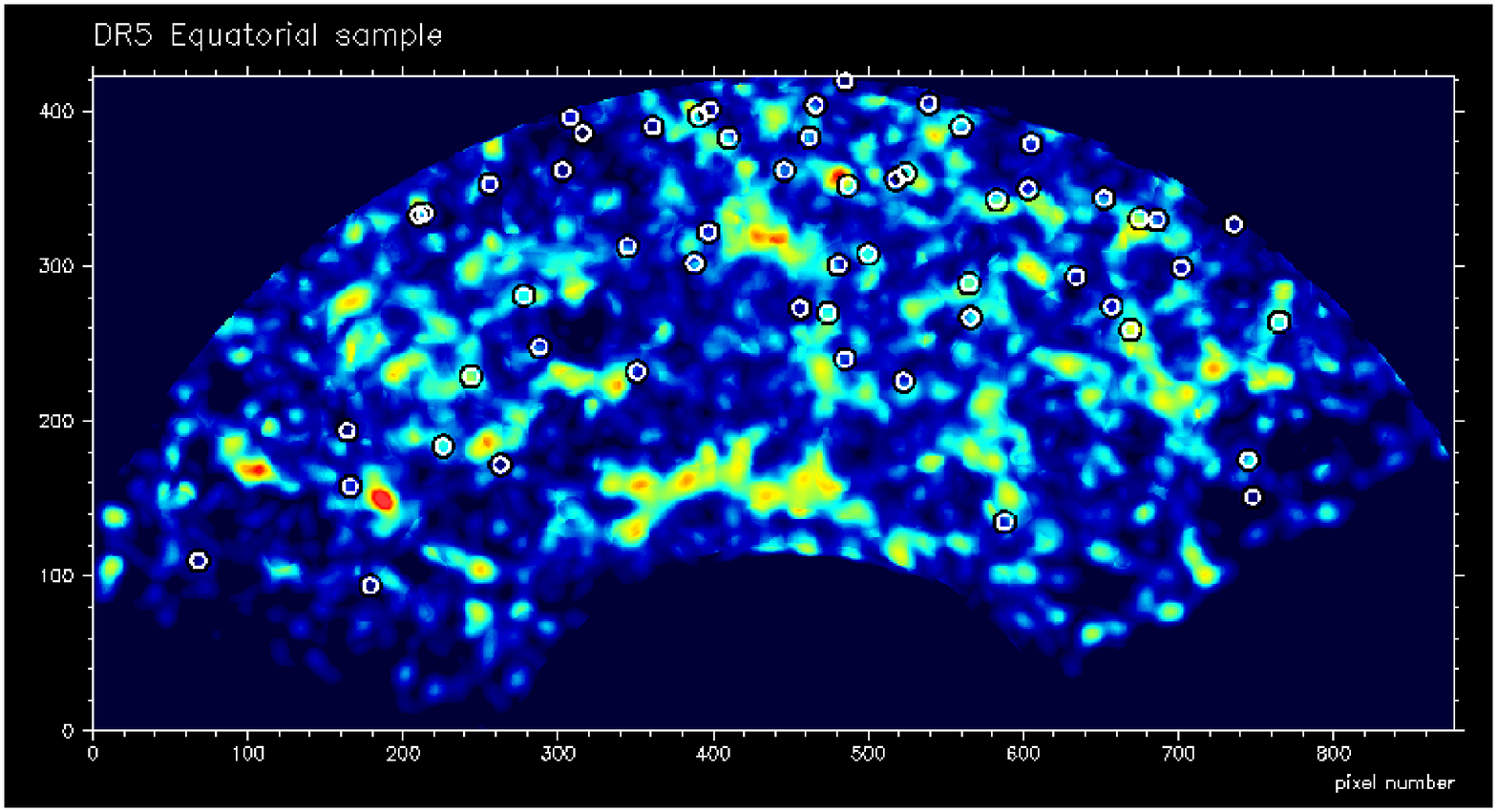}}
  \caption{Quasars in the luminosity density field of the northern sample (top) 
 and the equatorial sample (bottom) of the SDSS.}
  \label{ldenfield}
\end{figure*}

We assume that every galaxy is a visible 
member of a density enhancement (a group or a cluster). 
We correct the galaxy luminosities by a weighting factor that accounts 
for the expected group galaxies outside the visibility window:
\begin{equation}
 L_{\mathrm{tot}}=L_{\mathrm{obs}}W_L, 
\end{equation}
where 
$L_{\mathrm{obs}}$ is the luminosity of an observed galaxy and $W_L$ is the ratio of the 
total luminosity to the 
luminosity in the visibility window $L\in\left[L_1,L_2\right]$:
\begin{equation}
W_L =  {\frac{\int_0^\infty L \phi
(L)\mathrm{d}L}{\int_{L_1}^{L_2} L \phi (L)\mathrm{d}L}}\,;
\end{equation}
$\phi(L)$ is the 
galaxy luminosity function. 
According to Fig. 11 in Tago et al. (\cite{Tago}), the parameter $W_L$ 
is less than two at the distances from 100 to 300 $h^{-1}$Mpc, and rizes up to 
5 at 500 $h^{-1}$Mpc.

Finally, group and cluster galaxies are positioned at the mean group--cluster 
distance, to correct for the 
effect of the dynamical velocities (redshift space fingers). 

Next a Cartesian grid is defined in the survey volume and the luminosity 
density field on the grid is calculated using the B3-spline kernel function.
Each galaxy contributes its 
luminosity to the density field. 
This gives us the total luminosity density field of the survey where different characteristic structures have different density levels.

The total luminosity in a subvolume can be calculated by summing over all 
the grid vertices it contains (the luminosity density
units are $1.0\times10^{10}/(D^3) [L_{\sun}/($\Mpc$)^3]$, where $D$ is the grid cell 
size in Mpc/h). 
Using an appropriate mass-to-luminosity ratio one can 
translate luminosities into masses.
We chose the cell size of 1\,\Mpc, as
this is the characteristic size of a cluster. The half-width 
of the B3 smoothing kernel was 16\,\Mpc, and the effective 
smoothing radius (the kernel half-width for which the density value differs 
considerably from zero) is 
8\,\Mpc. This is a appropriate value to characterize the global environment 
and superclusters (Einasto et al. \cite{EinastoJ07}).

Regions of $D_L < 1.5$ times the mean density 
are considered to be voids. Superclusters are  
systems that occupy regions above the threshold density 
about $D_L > 4.6$ (Einasto et al. \cite{EinastoJ07}). 
This is approximately the lowest limit for separating 
non-percolating galaxy
systems (poor superclusters) from the overall density field. 
Densities which correspond to rich superclusters have a median value of 
$D_L=7.5$ times the mean density and $D_L >10$ correspond to cores of rich superclusters 
(Einasto et al. \cite{EinastoM07}).
These values were determined for the 2dFGRS data, and there may be 
some differences for the superclusters in the SDSS. However, the differences should 
not be large enough to influence our analysis.

Figure~\ref{ldenfield} shows the projected luminosity density field of the 
SDSS DR5 northern and equatorial samples. Red regions show the most luminous 
centers of the superclusters with densities of about 20 times the mean 
luminosity density. The circles in the figure represent quasars.   
As we see, visually quasars seem to be located in low density regions, in the outskirts of 
superclusters.

\begin{figure}[ht]
\centering
\resizebox{0.8\columnwidth}{!}{\includegraphics{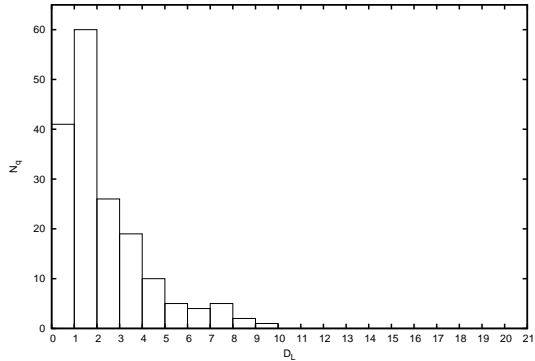}}
  \caption{The number of nearby quasars with a given
  environmental luminosity density $D_L$ (in a ball of the radius of
     2\,\Mpc.}
\label{nl}
\end{figure}

A more detailed analysis reveals that the local 
mean luminosity density around quasars is about 2.4 times the mean luminosity  $D_L$
(in units of mean luminosity density). This is illustrated in 
Fig.~\ref{nl} that shows the distribution of the 
mean luminosity density at quasar locations (in a ball of the radius
of  2\,\Mpc) -- the number of quasars with a given environmental
density.

Most of the quasars seem to be located in a low density 
environment on these scales.
Quasars  avoid densities which are characteristic for 
poor superclusters, $D_L > 4.6$,
and none is found in high density environments, $D_L > 10$.

\begin{figure*}[ht]
\centering
\resizebox{0.8\columnwidth}{!}{\includegraphics*{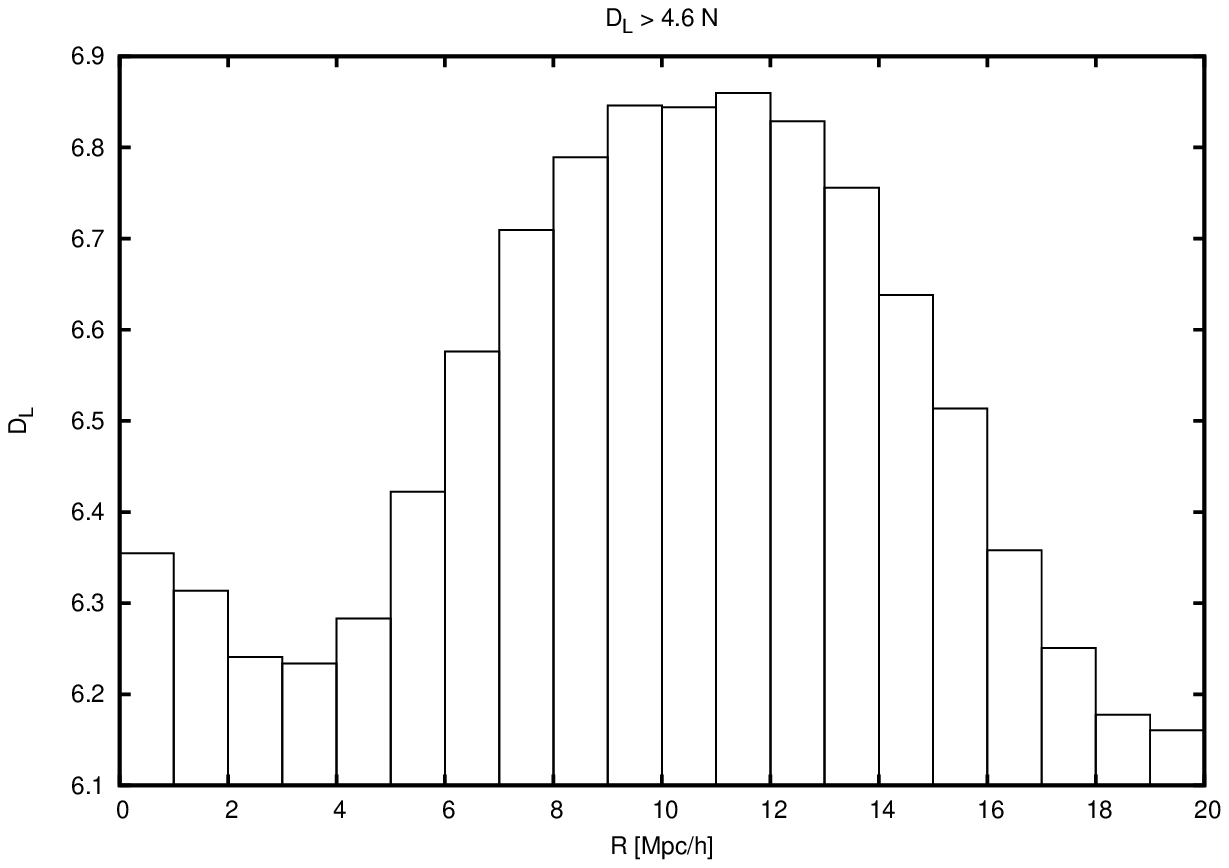}}
\resizebox{0.8\columnwidth}{!}{\includegraphics*{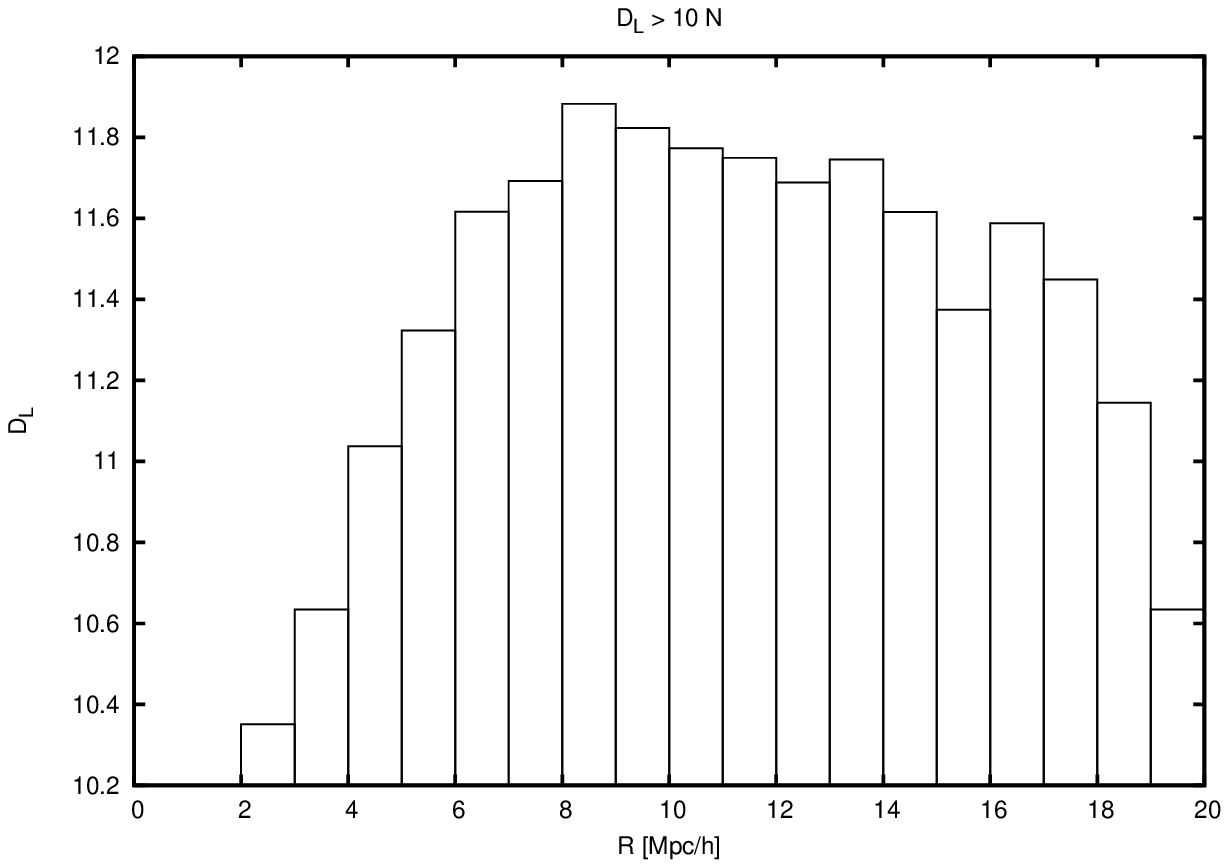}}
\hspace*{2mm}\\
\resizebox{0.8\columnwidth}{!}{\includegraphics*{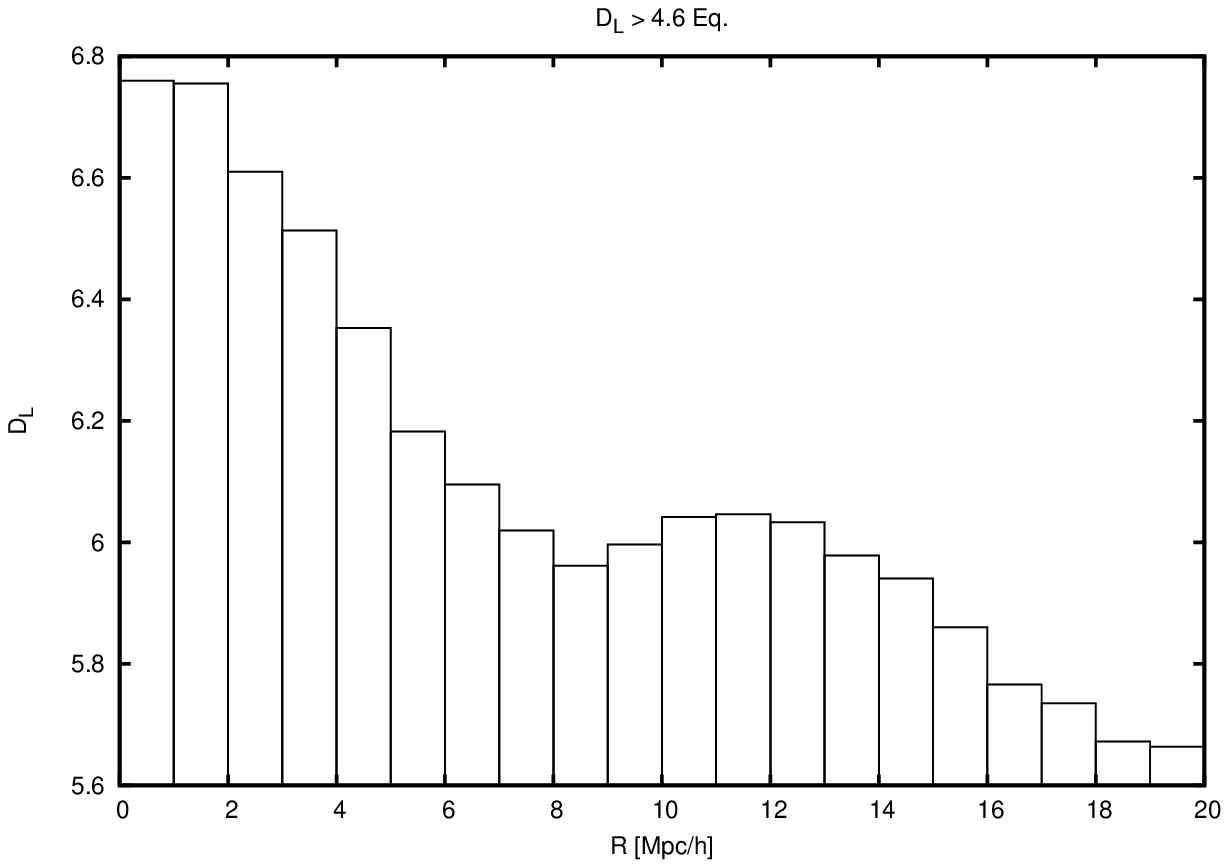}}
\resizebox{0.8\columnwidth}{!}{\includegraphics*{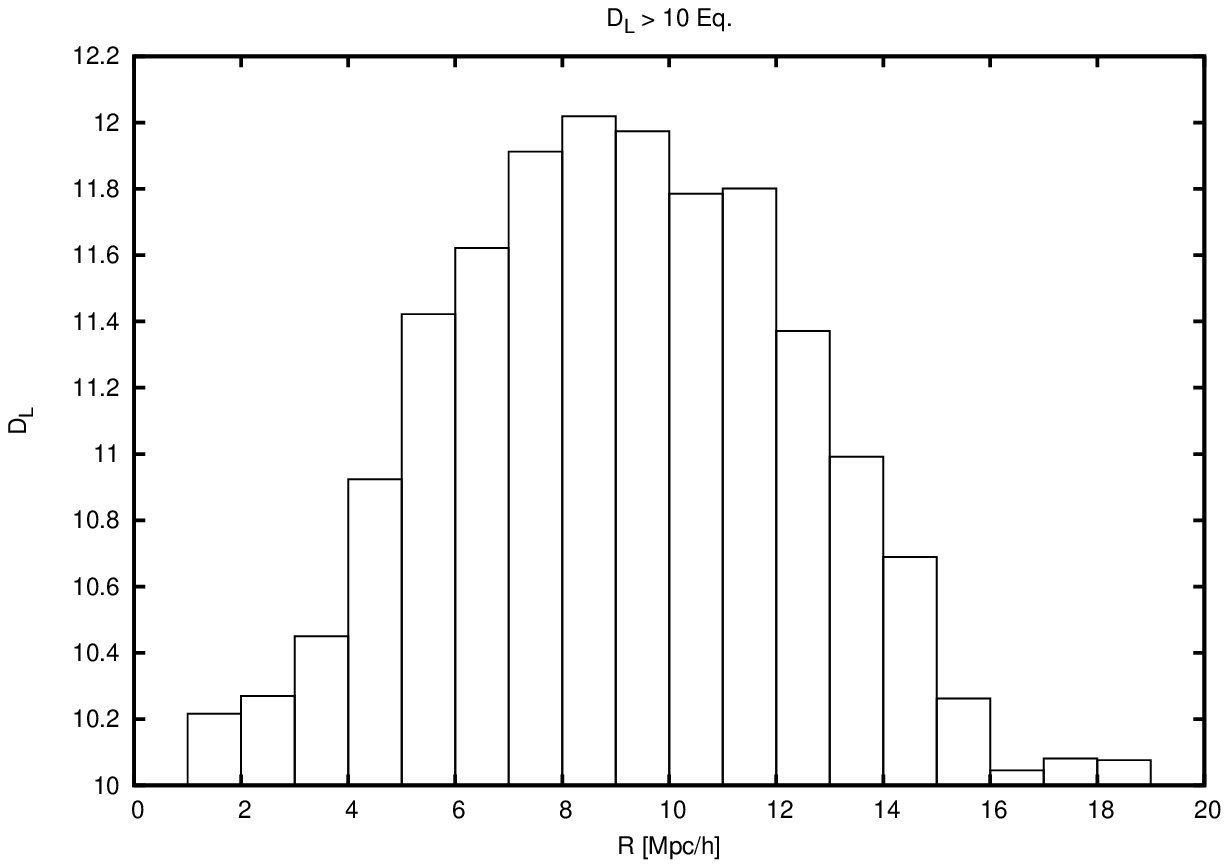}}

  \caption{ The mean luminosity density of galaxies in the north (N) and 
  equatorial (Eq.) SDSS-DR5 volumes as a function of distance from quasars.
The density thresholds are $D_L > 4.6$ and $D_L > 10$}.   
      \label{DLhis}
\end{figure*}

Figure~\ref{DLhis} shows the mean luminosity density of galaxies 
   in north and 
  equatorial samples
around quasars  as a function of
distance. Two different threshold densities, 
$D_L > 4.6$ and $D_L > 10$ are used to separate 
poor and rich superclusters.
As we see, regions with $D_L > 4.6$ belong to the nearby quasar environments. 
When the threshold value is  increased to $D_L > 10$, 
no nearby quasars are found.      
If we increase the threshold value even higher, the
high density regions are mostly located  
at the distance of approximately 10\,\Mpc 
from the quasars. 
This is close to the 
distance found in  the 
analysis based on the volume limited galaxy sample in section 3.2. 
This scale agrees with the characteristic scale for superclusters and
void sizes found in observations and simulations.
(von Benda-Beckmann et al. \cite{Benda},
Ceccarelli et al. \cite{Ceccarelli},
Gottl\"{o}ber et al. \cite{Gottlober}).

\begin{figure*}[ht]
\centering
\resizebox{0.8\columnwidth}{!}{\includegraphics*{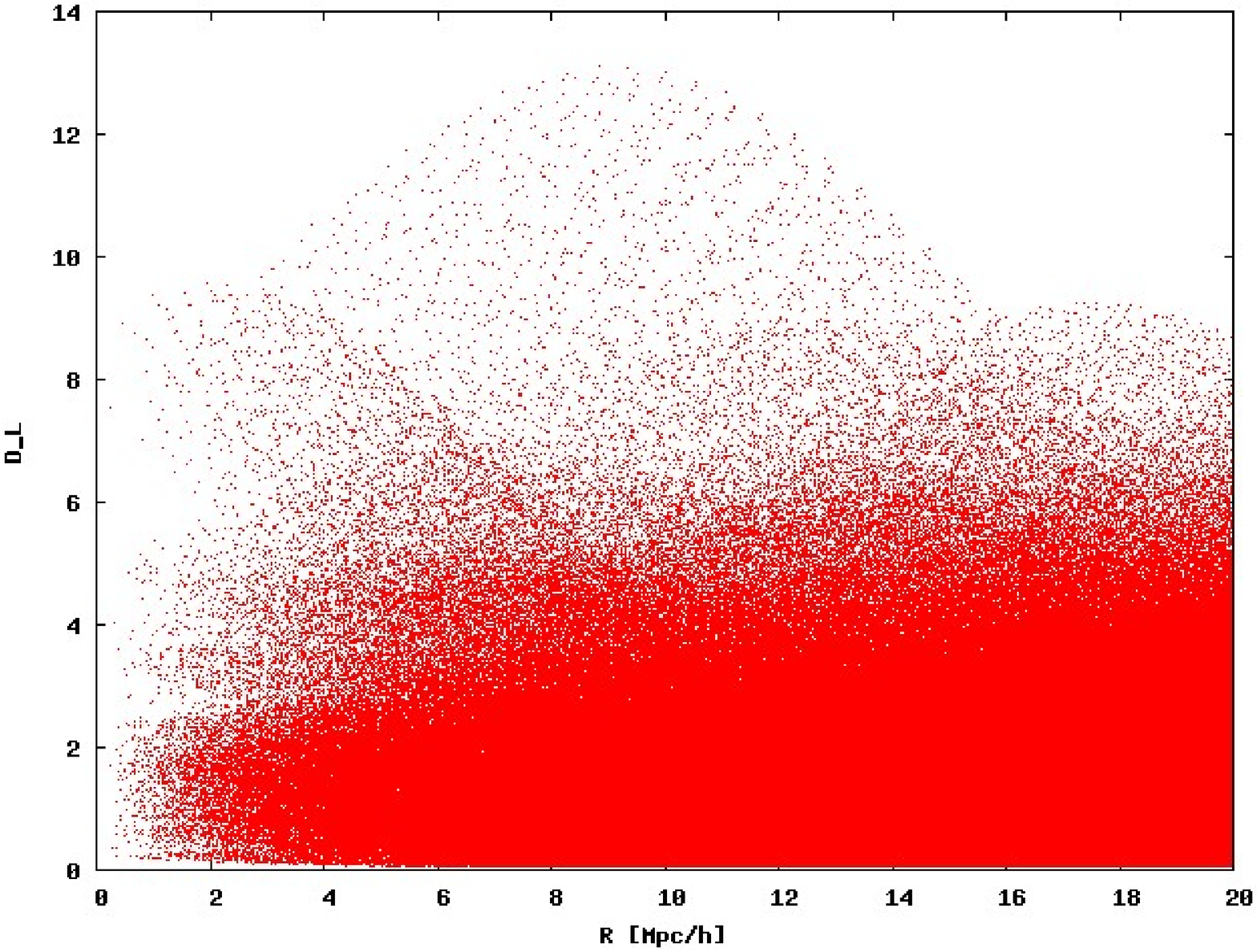}}
\resizebox{0.8\columnwidth}{!}{\includegraphics*{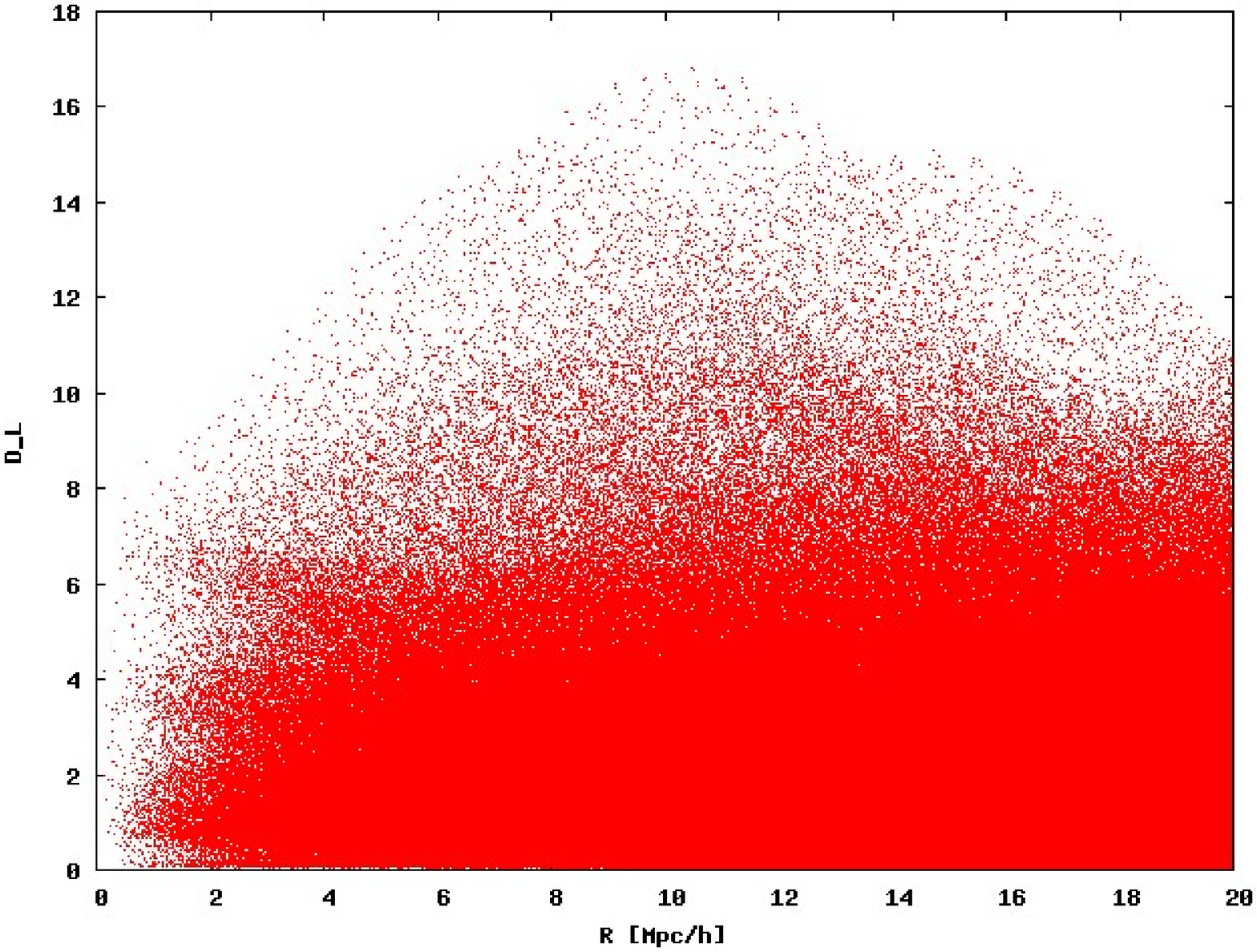}}
\caption{All luminosity density values in the north (right) and 
 equatorial (left) SDSS-DR5 volumes as a function of distance from quasars}.   
      \label{allLD}
\end{figure*}

This characteristic distance is clearly seen also in 
Fig.~\ref{allLD} which shows the $D_L$ values for the full 
luminosity density grid versus the distance to the nearest quasar.
In both panels of Fig.~\ref{allLD} we see the highest 
luminosity densities for the distances of about 10\,\Mpc from
quasars. 
Differences between northern and equatorial samples in Figs. 
\ref{DLhis} and \ref{allLD} are most probably due to quantitative and 
intrinsic variance of the relatively large and complex filamentary structures 
of the superclusters between two 
samples of different sizes. Volume of the equatorial sample is about half of 
the northern one. Differences between these samples are seen also in 
Fig. \ref{ldenfield}.
This analysis indicates that nearby quasars 
avoid high density regions 
and prefer poor superclusters, outskirts of rich superclusters and filaments.  

The most luminous galaxies as well as the richest groups are located in core 
regions of rich superclusters (Einasto et al. \cite{EinastoM07}). 
This is in accordance with the findings that quasars avoid rich groups 
and prefer filaments near superclusters.
Thus our luminosity density analysis agrees with our result in sections 3.1 and 3.2 where we 
showed that the galaxy number density around QSOs is a function of the 
luminosity of galaxies.

\section{Discussion}

One of our main results is that nearby quasars are located in low density 
environments among luminous galaxies on a few Mpc scale. 
This confirms the result by Coldwell \& Lambas (\cite{Coldwell}) 
for quasars in the same redshift range as here.
Strand et al. (\cite{Strand}) find that quasars are located in more dense regions 
compared to other types of active galaxies. They studied 
environments of quasars at scales $< 2$ Mpc, while we analysed mostly larger 
scales. Moreover, we studied only the distribution of bright galaxies.
It may be possible that the density of fainter galaxies in the close 
environments of quasars is higher.

There are several studies which demonstrate 
that the properties 
of galaxies and dark matter halos depend on their 
environments in several ways (Kauffmann et al. \cite{Kauffmann}, 
Harker et al.\cite{Harker} , Balogh et al. \cite{Balogh}, Tempel et al. \cite{Tempel}). 
For example, Constantin et al. (\cite{Constantin}) state that 
void and cluster galaxies follow different evolutionary paths, a point    
which should apply to active galactic nuclei as 
well. Presumably, the accretion process 
for the supermassive black hole (SBH) in a galaxy center also differs for 
distinct environments.

There is an observationally well justified relation for the masses of the 
SBH and the bulge velocity dispersion of the host 
galaxy (e.g. Magorrian et al. \cite{mag}, 
Graham et al. \cite{Graham}, Tremaine et al. \cite{Tremaine}).
McLure \& Dunlop (\cite{mc}) applied this relation for estimating the
BH masses for the 12698 quasars in the first release of the SDSS, and found that
quasar SBH  masses lie between $\sim 10^7 \Msun$ and $\sim 10^9 \Msun$. 
The connection between the growth of the black holes 
and the underlying structure formation 
was shown by  
Ferrarese (\cite{fer}). She pointed out 
that the mass of the SBH correlates 
with the mass of the dark matter halo in which 
they presumably were formed. 
Combining the results on quasar mass limits by McLure \& Dunlop (\cite{mc})  
with the results of Ferrarese (\cite{fer}) 
 (Fig.~5 and eq.~6 in her paper)  
we can estimate that the masses of the
dark matter host halos of the local quasars should 
have been around $10^{12}$ to $ 10^{13} \Msun$. 
These mass limits are compatible 
also with the result by Croom et al. (\cite{croom}). 
Moreover, the results by Croom et al. (\cite{croom}) suggest that the mass of the 
dark matter halo of a quasar is more or less independent of the redshift.  
This means that the present-day host  
dark matter halos of the nearby quasars should have the same
mass range $10^{12}-10^{13} \Msun$. 

Using simulations, Gao et al. (\cite{Gao}) 
showed that on large scales dark matter 
halo clustering depends on the age of the halo. 
They concluded that halos of given mass that assembled at high redshift 
are substantially more strongly clustered than halos of the same mass that 
assembled recently. The 'oldest' 10 per cent 
of halos are more than five times more strongly 
correlated than the 'youngest' 10 per cent.
Harker et al. (\cite{Harker}) studied the environmental 
dependence of halo formation times. They concluded that halos of a given 
mass in denser regions formed at higher redshifts than those in less 
dense regions. Maulbetsch et al. (\cite{Maul}) estimate that 
halos of $\sim 10^{12}\Msun$ assemble $\sim $1.5\,Gyr earlier 
in the high density 
regions than in the low density regions.
Using the Millennium simulation and arguments that link the
formation of the dark matter halos to the SBHs and to the formation of early 
quasars, Springel et al. (\cite{Spring}) also showed that the first quasars 
end up today as central galaxies in rich clusters.
As a result, nearby quasars in low density environments
should be younger than quasars in high density environments. 

Observationally the clustering of high redshift quasars in the Sloan Digital Sky Survey 
has been studied by Shen et al. (\cite{Shen}). Their results indicate  
increasing clustering for high redshift quasars with redshift.
Using the 2dFQSO Redshift Survey (2QZ) 
Croom et al. (\cite {croom}) showed that quasar clustering increases with redshift.
This suggests that earlier quasars were located in 
higher density environments than 
the recent ones, and high-redshift 
quasars can be used as beacons to search for 
high-density regions (Coldwell \& Lambas \cite{Coldwell}).
Our results of the location of the nearby quasars agree with 
these observations.  

An interesting question is also what is the relation between dark matter halos and 
their environment
with the observed redshift distribution 
of the number of quasars per unit volume. This distribution for the quasar main population
reaches a maximum around  $z \simeq 2-3$ both in X-ray selected and 
optically selected surveys (Silverman et al.\cite{Silver}).

Maulbetsch et al. (\cite{Maul}) address the problem how to explain the
observed dependencies of galaxies on the
environment (the morphology--density relation) with the
mass assembly history of dark matter halos.  
They show that aggregation of mass to the dark matter halos via  
halo mergers and mass accumulation has a strong dependence on the 
environment and redshift.  
While halos in low density environments grow mainly  
by  mass accumulation,
in high density environments halo mergers are the dominant aggregation 
procedure. The latter one is mainly determined by the properties of the subhalo population  
and the survival of subhalos in the main dark matter halo. 
The mass fraction accumulated by mergers is a function of redshift and 
it decreases towards $z=0$. This concerns both the low and high density environments,
but the decrease is steeper in low density regions. At the same time, at $z=0$   
mergers of small mass halos 
are relatively more numerous in low density environments.  
This is interesting since we know that
quasar activity is usually expected to be triggered by galaxy interactions.

Finally, for supercluster scales,
numerical simulations by Einasto et al. (\cite{EinastoJ05})
showed that in voids the dynamical evolution of halos is very slow and 
finally stops, while in supercluster regions the dynamical evolution 
continues until the present day. 
In superclusters, rich clusters in the core regions form first via halo mergers. In less 
dense regions of the superclusters and rich filaments the
halo evolution is suppressed compared to the core regions. 
  
Thus it is not a far-fetched idea that 
dark matter halos, evolution of the mass 
aggregation rate and its dependence on environment 
on all scales play an 
important role in quasar evolution and their observed redshift distribution.
 The 
triggering mechanisms,
which influence aggregation onto black holes, may 
 have been 
suppressed due to environmental reasons, which are also related to the 
evolution of superclusters. 
Of course,  local astrophysical 
processes related to baryonic matter 
must explain the observed properties of quasars in detail.

Taking into account halo formation times 
and merger rates as a 
function of the halo mass 
and halo environment, also on supercluster scales, together with the estimated quasar 
duty-cycle, it should be possible to generate quasars in cosmological simulations,
and to compare the results with the observed quasar distribution.
We are planning to do this in our forthcoming paper.

\section{Conclusions}

We have studied the environments of quasars in the SDSS on different scales. 
Our main results can be summarized as follows:

(i) At the distances from 2 to 20\,\Mpc from quasars the number density of 
galaxies is lower than the density at the same distances from galaxies. 
This effect is particularly strong for the brightest 
galaxies.

(ii) The groups of galaxies that have a quasar closer than 2\,\Mpc 
are poorer and less luminous than all the groups in average. Also,
groups that consist of four or more galaxies have lower richnesses and 
luminosities if there is a quasar closer than 10\,$h^{-1}$Mpc.

(iii) The groups of galaxies that have a quasar neighbour at a 10 to 15 Mpc distance are 
richer and more luminous than the groups in average. The same result can 
also be seen if we examine the rich groups only.

(iv) Quasars avoid the densest regions in the (luminosity) density field, which 
correspond to rich supercluster cores. Nearby quasars are located 
in the outskirts of superclusters or in filaments connecting them. Quasar evolution 
may be related to these largest density enhancements in the Universe.

\begin{acknowledgements}
We thank the anonymous referee for constructive comments that helped us 
improve the paper

This project was supported by the Finnish Academy funding.
H. Lietzen was supported by Magnus Ehrnrooth foundation. 
E.Tago, E. Saar, J. Liivam\"agi, E. Tempel, M. Einasto, J. Einasto and M. 
Gramann were  supported by the Estonian Science Foundation grant No. 6104 and 
7146, and by the Estonian Ministry for Education and Science research project 
TO 0060058S98.

Funding for the SDSS and SDSS-II has been provided by the Alfred P. Sloan 
Foundation, the Participating Institutions, the National Science 
Foundation, the U.S. Department of 
Energy, the National Aeronautics and Space Administration, the 
Japanese Monbukagakusho, the Max Planck Society, and the Higher Education 
Funding Council for England. The SDSS Web Site is http://www.sdss.org/.

The SDSS is managed by the Astrophysical Research Consortium for the 
Participating Institutions. The Participating Institutions are the 
American Museum of Natural History, Astrophysical Institute Potsdam, 
University of Basel, University of Cambridge, Case Western Reserve 
University, University of Chicago, 
Drexel University, Fermilab, the Institute for 
Advanced Study, the Japan Participation Group, Johns 
Hopkins University, the Joint Institute for Nuclear 
Astrophysics, the Kavli Institute for Particle Astrophysics 
and Cosmology, the Korean Scientist Group, the 
Chinese Academy of Sciences (LAMOST), Los Alamos 
National Laboratory, the Max-Planck-Institute 
for Astronomy (MPIA), the Max-Planck-Institute 
for Astrophysics (MPA), New Mexico State 
University, Ohio State University, University of 
Pittsburgh, University of Portsmouth, Princeton University, 
the United States Naval Observatory, and the University of Washington.

\end{acknowledgements}

\end{document}